\documentclass[reprint,amsmath,amssymb,aps,prd]{revtex4-2}

\usepackage{graphicx}%
\usepackage{dcolumn}%
\usepackage{bm}%
\usepackage{hyperref}%

\usepackage{aas_macros}%

\usepackage{soul} %
\usepackage[dvipsnames]{xcolor}
\usepackage{subfig}
\usepackage{ulem}  %

\begin{document}

\title{How accurate is Spectral Einstein Code (SpEC)?\texorpdfstring{\\}{ }And what limits its accuracy?}

\author{Himanshu Chaudhary}
\email{hchaudha@caltech.edu}
\affiliation{Theoretical Astrophysics 350-17, California Institute of Technology, Pasadena, CA, 91125, USA}%
\author{Mark A. Scheel}
\affiliation{Theoretical Astrophysics 350-17, California Institute of Technology, Pasadena, CA, 91125, USA}
\author{Robert Owen}
\affiliation{Department of Physics and Astronomy, Oberlin College}
\author{Saul A. Teukolsky}
\affiliation{Theoretical Astrophysics 350-17, California Institute of Technology, Pasadena, CA, 91125, USA}
\affiliation{Cornell Center for Astrophysics and Planetary Science, Cornell University, Ithaca, New York 14853, USA}
\author{Michael Boyle}
\affiliation{Cornell Center for Astrophysics and Planetary Science, Cornell University, Ithaca, New York 14853, USA}
\author{Lawrence E. Kidder}
\affiliation{Cornell Center for Astrophysics and Planetary Science, Cornell University, Ithaca, New York 14853, USA}
\author{Harald P. Pfeiffer}
\affiliation{Max Planck Institute for Gravitational Physics (Albert Einstein Institute), Am M{\"u}hlenberg 1, 14476 Potsdam, Germany}

\collaboration{SXS Collaboration}%

\date{October 16, 2026}%

\begin{abstract}
The improved sensitivity of upcoming gravitational-wave detectors will require correspondingly more accurate binary black hole (BBH) simulations. Given the high cost of such simulations, it is essential to understand their sources of error and use computational resources efficiently. We study sources of error in BBH simulations performed with the Spectral Einstein Code (SpEC) and seek to understand what limits their accuracy. In particular, we focus on the inspiral phase because it is the longest-running part of the simulation and accounts for the majority of the waveform error. We find that targeted changes to simulation parameters and resolution in different parts of the computational domain reduce constraint violations and gravitational waveform differences. We also discuss more fundamental limitations on our accuracy that become important at higher resolutions: floating-point errors and large differences in the scales of the fundamental variables. Addressing these limitations will require substantial changes to both the SpEC code and the formulation it uses to evolve Einstein’s equations. Many of the improvements and changes discussed in this paper are likely to be relevant to other numerical relativity codes.

\end{abstract}

\maketitle

\section{Introduction}
\label{sec:Introduction}

The LIGO–Virgo–KAGRA Collaboration \cite{sxs25_5,sxs25_6} has detected hundreds of binary black hole (BBH) mergers since the first detection in 2015 \cite{sxs25_1,sxs25_2,sxs25_7,sxs25_8,sxs25_10,LIGOScientific:2025slb,LIGOScientific:2026wfs}. These observations have been used to study the evolution of binary black hole systems and their population distribution in our universe \cite{sxs25_9,LIGOScientific:2025pvj,LIGOScientific:2026ctl}. They also provide some of the most accurate tests of General Relativity \cite{TestsOfGR_Ynues_LRR_2025,LIGOScientific:2025wao,LIGOScientific:2026oim}, by allowing us to probe the highly non-linear regimes of the theory. The next generation gravitational wave detectors, like the Laser Interferometer Space Antenna (LISA) \cite{sxs25_78,LISA:2024hlh}, Cosmic Explorer (CE) \cite{sxs25_74}, Einstein Telescope (ET) \cite{sxs25_76,sxs25_77} and LIGO A$^\sharp$ \cite{LIGO:2026gzb} will further increase the number and quality of gravitational wave signals.

Analyzing data from GW detectors requires accurate models of the gravitational-wave signals produced by BBH mergers, both to search for events and to infer source parameters such as the black hole masses and spins. The only {\it ab initio} method of producing such models for inspiral, merger, and ringdown is numerical relativity: the direct solution of Einstein's equations on the computer.

The high computational cost of numerically evolving binary black hole mergers means that numerical relativity simulations are usually not used directly for data analysis. Instead, they are used to construct surrogate models \cite{sxs25_68,sxs25_71,sxs25_72,sxs25_69,sxs25_70}, which interpolate gravitational waveforms to points in parameter space not directly covered by numerical simulations. They are also used as the ground truth for calibrating other models like Effective-One-Body (EOB) models \cite{sxs25_43,sxs25_50,Pompili:2023tna,Gamboa:2024hli,Gamboa:2026ybr} and phenomenological models \cite{Pratten:2020ceb,Colleoni:2024knd,Thompson:2023ase,Hamilton:2025xru}.

Given the importance of numerical relativity waveforms, it is important that the simulations are accurate enough to avoid systematic biases, especially for the higher-sensitivity of the next generation of detectors. According to some estimates \cite{sxs25_83,sxs25_84}, waveform mismatches need to be reduced by at least an order of magnitude to be sufficiently accurate for next-generation GW detectors. Many new codes that leverage newer hardware and scale better are being developed to simulate these systems at much higher resolutions, e.g., \cite{spectrecode,AthenaK_Jim_Stone:2026,ETK_CarpetX:2025}. In this paper, we will focus on the NR code SpEC, which was used for the simulations in the SXS catalog \cite{Scheel:2025jct}. SpEC is currently one of the most accurate and efficient codes for BBH simulations \cite{Scheel:2025jct}. Nevertheless, a large fraction of the waveforms in the catalog do not meet these accuracy requirements. The primary goal of this paper is to address the question: How accurately can we evolve a BBH in SpEC if we had infinite computing resources and time? If the only limiting factor is compute time, then the new generation codes that scale better will allow us to reach these accuracy requirements. On the other hand, if there are more fundamental issues in the formalism or the numerical algorithms, then we need to resolve those before we do more high-resolution simulations. For this study, we use simulations of an equal-mass, non-spinning BBH system during the inspiral.

This paper is organized as follows: In Section~\ref{sec:Simulation Setup and Spec Background}, we briefly describe the simulation setup and terminology used in the later sections. In Section~\ref{sec:Accuracy metrics in SpEC}, we describe constraint violations and waveform differences, the two main quantities used to quantify the accuracy of SpEC simulations. In Section~\ref{sec:Improvements to SpEC}, we discuss the changes we made to SpEC and how they improve these accuracy metrics. Finally, in Section~\ref{sec:What limits our accuracy?}, we discuss factors that limit our accuracy and potential future improvements.

Throughout the paper, we will use Latin letters $a$, $b$, \dots for spacetime indices and $i$, $j$, $k$, \dots for spatial indices. We use the metric signature $(-,+,+,+)$ and set $G=c=1$.

\section{Simulation Setup and Spec Background}
\label{sec:Simulation Setup and Spec Background}

\subsection{Generalized Harmonic System}
\label{ssec:Generalized Harmonic System}

To solve Einstein's equations numerically, it is useful to recast them into first-order hyperbolic form, which allows us to establish a well-posed initial-value problem and use well-developed numerical algorithms. This can be done in multiple ways \cite{NR_Baumgarte, NR_shibata}, and SpEC uses the Generalized Harmonic (GH) formalism \cite{sxs25_150,sxs25_151,sxs25_152,sxs25_153}. The starting point of the GH formalism is the identity
\begin{align}
    \begin{aligned}
        R_{a b}&=-\frac{1}{2} g^{c d} \partial_c \partial_d g_{a b}+\nabla_{(a} \Gamma_{b)} \\
        &+g^{c d} g^{e f}\left(\partial_e g_{c a} \partial_f g_{d b}-\Gamma_{a c e} \Gamma_{b d f}\right),
    \end{aligned}
    \label{eqn:Ricci_in_terms_of_g_and_gamma}
\end{align}
where $g_{ab}$ is the spacetime metric and $\Gamma_a\equiv g^{b c} \Gamma_{a b c}$ is the trace of the Christoﬀel symbol. To fix the coordinate gauge freedom, we choose a gauge source function $H^a$ and require the coordinates $x^a$ to satisfy
\begin{equation}
    g_{a b} \nabla^c \nabla_c x^b=H_a.
    \label{eqn:H_in_terms_of_coordinates}
\end{equation}
From Eq.~\eqref{eqn:H_in_terms_of_coordinates}, we can see that by fixing $H_a$, along with appropriate initial and boundary conditions, we can fix all the coordinate gauge freedoms. Because of the identity
\begin{equation}
    g_{a b} \nabla^c \nabla_c x^b=-\Gamma_a,
\end{equation}
the gauge source function $H_a$ is related to the trace of the Christoﬀel symbol $\Gamma_a$ as $H_a = -\Gamma_a$. Thus, the Ricci tensor can be written as
\begin{align}
    \begin{aligned}
    R_{a b}&=-\frac{1}{2} g^{c d} \partial_c \partial_d g_{a b}-\nabla_{(a} H_{b)} \\
    &+g^{c d} g^{e f}\left(\partial_e g_{c a} \partial_f g_{d b}-\Gamma_{a c e} \Gamma_{b d f}\right).
    \end{aligned}
\end{align}
We write the spacetime metric in $3+1$ form as
\begin{equation}
    g_{a b}\,\mathrm{d}x^a\mathrm{d}x^b
    =-N^2\mathrm{d}t^2+\gamma_{i j}
    \left(\mathrm{d}x^i+N^i\mathrm{d}t\right)
    \left(\mathrm{d}x^j+N^j\mathrm{d}t\right),
\end{equation}
where $N$ is the lapse, $N^i$ is the shift, and $\gamma_{ij}$ is the spatial metric, with inverse $\gamma^{ij}$. The future-directed timelike unit normal to the constant-$t$ hypersurfaces is
\begin{equation}
    t^a=\frac{1}{N}\left[(\partial_t)^a-N^i(\partial_i)^a\right].
\end{equation}
To convert the GH equations into first-order form, we define two new variables $\Pi_{ab}$ and $\Phi_{iab}$:
\begin{align}
\begin{aligned}
N \Pi_{a b}&\equiv-Nt^c\partial_c g_{a b}
    =-\partial_t g_{a b}+N^i\Phi_{i a b}, \\
\Phi_{i a b}&\equiv\partial_i g_{a b}.
\end{aligned}
\end{align}
Using these, we can write down the principal part of the GH system,
\begin{align}
    \begin{aligned}
    \partial_t g_{a b}-\left(1+\gamma_1\right) N^k \partial_k g_{a b} &\simeq 0 \\
    \partial_t \Pi_{a b}-N^k \partial_k \Pi_{a b}+N \gamma^{k i} \partial_k \Phi_{i a b}& \\
    -\gamma_1 \gamma_2 N^k \partial_k g_{a b} & \simeq 0 \\
    \partial_t \Phi_{i a b}-N^k \partial_k \Phi_{i a b}+N \partial_i \Pi_{a b}-\gamma_2 N \partial_i g_{a b} & \simeq 0 .
    \label{eq_principal_parts_GH_system}
    \end{aligned}
\end{align}
Here, $\simeq$ means that we have omitted lower-order terms that do not contain derivatives of the evolved fields. The quantities $\gamma_1$ and $\gamma_2$ are free parameters. The parameter $\gamma_1$ modifies the coordinate characteristic speed, while $\gamma_2$ is used for constraint damping \cite{sxs25_153}. Constraint damping is discussed further in Section~\ref{ssec:Constraint violations and Constraint damping}. Full equations for the GH system are given in Appendix~\ref{app:Generalized Harmonic System}. In SpEC, $\Pi_{ab}$ and $\Phi_{iab}$ are combined into a single variable called $\kappa_{abc}$,
\begin{align}
\begin{aligned}
& \kappa_{ibc} = \partial_{i} g_{bc} \\
& \kappa_{0bc} = \Pi_{bc}.
\end{aligned}
\end{align}
Together, $\kappa_{abc}$ and $g_{ab}$ form the set of fundamental variables in our first-order GH system. Variables $H_a$, $g_{ab}$, and $\kappa_{abc}$ are not independent of each other and should satisfy constraint equations, which will be discussed later in Section~\ref{ssec:Constraint violations and Constraint damping}.

\subsection{Computational domain in SpEC}
\label{ssec:Computational domain in SpEC}

\begin{figure}
  \centering
  \includegraphics[width=0.48\textwidth]{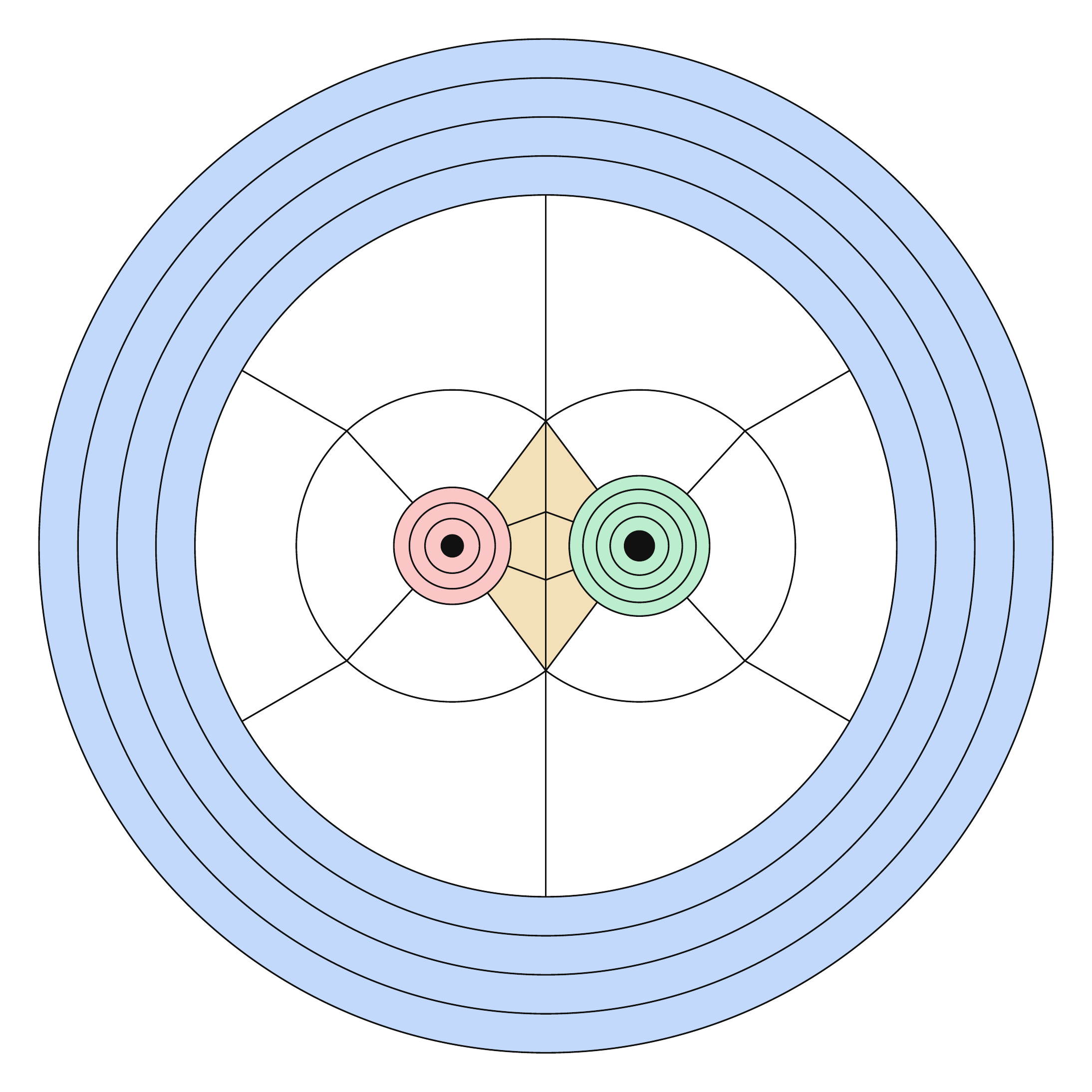}
  \caption{The $z=0$ cross-section of the subdomains used in SpEC for simulating a generic binary black hole system. Spherical shell subdomains in green are the SphereAs, with SphereA0 being the innermost subdomain that contains the horizon of black hole A. Similarly, red subdomains are SphereBs, and they contain the black hole B. The outermost spherical shell subdomains in blue are SphereCs, with SphereC0 being the smallest of these. The remaining subdomains are deformed cylindrical shells and deformed cylinders that join the inner SphereA and SphereB subdomains with the outer SphereC subdomains. Those connecting the SphereA and SphereB sets are highlighted in pale gold. These subdomains are further divided into smaller subdomains by the adaptive mesh refinement (AMR) algorithm as required. Note that the figure is not to scale, and in general, a typical SphereC subdomain is an order of magnitude wider than a typical SphereA and SphereB subdomain.}
  \label{fig_SpEC_subdomains}
\end{figure}

To solve Einstein's equations, we decompose the 3D volume into a set of simple shapes that fill the volume without overlapping: spherical shells, deformed cylindrical shells, and deformed cylinders. These simple shapes are called subdomains, and Figure~\ref{fig_SpEC_subdomains} shows in a 2D cross-section how these subdomains are joined together. This computational domain is primarily designed to handle the two excised regions that remove the black hole singularities \cite{sxs25_157,sxs25_159,sxs25_160,sxs25_162}. There are spherical shells around each excision surface, called SphereAn and SphereBn, where $n$ is an integer. SphereA0 is the innermost subdomain which contains the apparent horizon of the larger black hole, and it is surrounded by SphereA1, and so on. Similarly, in the wave zone, we have a set of spherical subdomains denoted by SphereCn, with SphereC0 being the innermost spherical subdomain. The innermost SphereC, the outermost SphereA, and the outermost SphereB subdomains are joined using cylindrical subdomains oriented along the axis connecting the centers of the holes, and distorted to fill the space in between. The handling of boundary conditions at inter-domain boundaries and at the outer boundary is described in \cite{sxs25_153,sxs25_163,sxs25_164,sxs25_166}. More information about the domain structure used in SpEC can be found in Section~3 of \cite{Szil_gyi_2014} and the appendix of \cite{Buchman_2012}. This paper will primarily use diagnostics from SphereA0, which contains one of the apparent horizons, and from SphereC6, one of the subdomains from which gravitational waveform data is extracted.

\subsection{Pseudo-Spectral Methods and Power Monitors}
\label{ssec:Pseudo-Spectral Methods and Power Monitors}

In each subdomain, we use pseudo-spectral methods \cite{2001cfsm.book.....B,trefethen_ATAP_2019,Szil_gyi_2014} to write the evolved variables as a sum of (time-independent) spatial basis functions multiplied by (time-dependent) spectral coefficients. For example, in a spherical shell, the basis functions are Chebyshev polynomials in radius and spherical harmonics in angles. We take spatial derivatives by differentiating the basis functions, which we can do exactly; the approximation is that we use only a finite number of basis functions, and the ``resolution'' of the simulation corresponds to how many basis functions we retain. In this method, the set of time-dependent GH evolution equations can be thought of as a set of complicated coupled Ordinary Differential Equations (ODEs) in time for the spectral coefficients.

With each choice of basis functions, there is a particular choice of collocation points on the spatial grid that can be used to evaluate variables using Lagrange interpolation. The advantage of having both representations is that we can evaluate nonlinear terms easily in physical space, while derivatives are evaluated with spectral accuracy. In the actual code, it is the values at the collocation points in physical space that are the primary variables, but conceptually it can be more helpful to think in terms of the spectral coefficients. The error analysis in this paper will in fact focus on the spectral coefficients. As one might expect, we can go back and forth between the vectors of values at the collocation points and spectral coefficients by matrix multiplication.

To solve the ODEs, we use the Dormand--Prince 5(4) method \cite{Dormand_Prince_Ode_methods_1980}, which is an ODE solver from the Runge-Kutta family. This is an adaptive time-stepping method that is fifth-order accurate, using an embedded fourth-order estimate to compute the local error. This local error is used to change the step size adaptively and keep the error in our solution below the chosen ODE tolerance. SpEC simulations done till now have used a fixed ODE tolerance of $10^{-8}$. For typical simulations, the error is dominated by spatial truncation error and not temporal truncation error, so for a wide range of ODE tolerances, the adaptive timestepper tends to automatically settle to the Courant--Friedrichs--Lewy (CFL) limit: taking a timestep slightly larger than the CFL limit induces an instability, momentarily increasing the error until the timestepper responds by decreasing the timestep again. We will see in Section~\ref{ssec:Adaptive ODE Tolerance for Near-Horizon Constraint Convergence} that these occasional brief CFL violations cause noise in the solution.

The spectral coefficients ($C_{k_1 k_2 k_3}$) of a single tensor component form a three-index object, with each index associated with one coordinate direction of the associated three-dimensional subdomain. For example, in spherical-shell subdomains such as SphereA0, the spectral coefficients can be written as $C_{k\ell m}(t)$, where $k$ labels the radial Chebyshev modes and $(\ell,m)$ label the spherical-harmonic modes. Following Ref.~\cite{Szil_gyi_2014}, we define the radial power monitors by
\begin{align}
P^{(r)}_k(t)
  = \sqrt{\frac{1}{N_{\mathrm{ang}}}
    \sum_{\ell,m}\left|C_{k\ell m}(t)\right|^2},
\end{align}
where the sum runs over the $N_{\mathrm{ang}}$ retained angular modes. The superscript $(r)$ denotes the radial spectral basis, and there is one power monitor for each radial mode $k=0,\ldots,N_r-1$. Thus, $P^{(r)}_k$ is the root-mean-square magnitude of the spectral coefficients associated with radial mode $k$. For tensor fields such as $g_{ab}$ and $\kappa_{abc}$, the power monitors plotted in this paper additionally include a root mean square (RMS) over the stored independent tensor components. Power monitors for the other spectral bases are defined analogously.

These power monitors are used to estimate the relative truncation error,
\begin{align}
\begin{aligned}
T \left[P^{(r)}\right] & \equiv \frac{\prod_{k=0}^{N_r-1} \left(P^{(r)}_k\right)^{w_k / \sum_{q=0}^{N_r-1} w_q}}{\max \left(P^{(r)}_0, P^{(r)}_1\right)} \\
w_k & \equiv \exp \left(-\left(k-N_r+\frac{3}{2}\right)^2\right).
\label{eqn:Truncation Error}
\end{aligned}
\end{align}
where $N_r$ is the number of radial modes. The numerator is a weighted geometric mean of the power monitors, with most of the weight shared equally by the last two modes. We divide this by the larger of the first two power monitors to get a relative error estimate. This differs from the simpler definition of the truncation error, where the numerator is just the last mode and the denominator is the first mode. We use this weighted average to make the estimate vary more smoothly as the power monitors change, helping to make the decisions taken by the AMR less erratic~\cite{Szil_gyi_2014}. Analogous truncation-error estimates can be constructed for the other spectral bases. Roughly speaking, the truncation error estimates the error introduced by representing a smooth function using only a finite number of basis functions. The AMR algorithm computes the corresponding truncation error separately for each spectral basis of each subdomain and uses these estimates to control the grid resolution, which we discuss next.

\subsection{AMR tolerance and definition of levels}
\label{ssec:AMR tolerance and definition of levels}

We need a good AMR algorithm for a BBH simulation because the gravitational field changes rapidly near the black holes and is comparatively smoother in the wave zone. The AMR algorithm tries to change the grid resolution so that the truncation errors of the two fundamental variables ($g_{ab}$ and $\kappa_{abc}$) stay below the specified AMR tolerance. A smaller AMR tolerance generally requires more basis functions and results in a higher grid resolution. To study the numerical convergence of a simulation, we evolve the same initial data multiple times, but with different AMR tolerances. We parametrize the AMR tolerance of a simulation by a single integer, called its level. The base-tolerance is given by
\begin{align}
\begin{aligned}
    \epsilon_{\mathrm{base}}
    &= 10^{-4}\left(\frac{4}{e}\right)^2 4^{-\text{level}} \\
    &\approx 2.16536\times10^{-4}\times4^{-\text{level}}.
\end{aligned}
\end{align}
Here, $e$ is Euler's number. This means that if the levels of two resolutions differ by 1, then their base AMR tolerance differs by a factor of 4. To describe the spatial dependence of the tolerance, we define the squared Euclidean coordinate distances by
\begin{equation}
\begin{aligned}
r_A^2 &= |\mathbf{x}-\mathbf{x}_A|^2, \\
r_B^2 &= |\mathbf{x}-\mathbf{x}_B|^2, \\
r^2 &= |\mathbf{x}|^2,
\end{aligned}
\label{eq:coordinate_distances}
\end{equation}
where $\mathbf{x}_A$ and $\mathbf{x}_B$ are the coordinate centers of the two black holes \cite{sxs25_160}. We define the Gaussian weights
\begin{equation}
    w_A=e^{-r_A^2/\lVert\mathbf{x}_B\rVert^2},\qquad
    w_B=e^{-r_B^2/\lVert\mathbf{x}_A\rVert^2}.
\end{equation}
The default spatially varying tolerance is then
\begin{equation}
    \epsilon_{\mathrm{AMR}}(\mathbf{x})
    =\epsilon_{\mathrm{base}}\,10^{-4(w_A+w_B)/(1+w_A+w_B)}.
    \label{eq:spatial_amr_tolerance}
\end{equation}
Near either center, its Gaussian weight is approximately one and the other is small, giving $\epsilon_{\mathrm{AMR}}\approx\epsilon_{\mathrm{base}}/100$. Far from both holes, the weights approach zero and the tolerance approaches $\epsilon_{\mathrm{base}}$.

Later in Section~\ref{sec:Improvements to SpEC} we will also look at spatially uniform tolerance,
\begin{equation}
    \epsilon_{\mathrm{AMR}}=\frac{\epsilon_{\mathrm{base}}}{100},
    \label{eq:uniform_amr_tolerance}
\end{equation}
This leaves the near-hole tolerance approximately unchanged while making the outer tolerance a factor of 100 smaller. Furthermore, the level of a simulation also sets a minimum number of grid points for each subdomain. Finally, if the constraints are too large in a subdomain, then the AMR algorithm decreases the local AMR tolerance, thereby increasing the resolution of that subdomain.

\subsection{Junk radiation}
\label{ssec:Junk radiation}

An unsolved problem in numerical relativity is how to construct BBH initial data that truly corresponds to a snapshot of a BBH that has been inspiraling since the infinite past.  The actual BBH initial data we use are constraint-satisfying, but they have different horizon shapes and different (typically zero) gravitational waves compared to a true snapshot. As a result, evolving BBH initial data produces a transient high-frequency pulse of gravitational radiation, commonly referred to as ``junk radiation'' \cite{JunkRadiation_Lovelace_2009}. Junk radiation would be computationally expensive to resolve. However, because it is not part of the solution we are looking for, SpEC simulations do not attempt to resolve it. As a result, the unresolved junk radiation produces large constraint violations.  The early part of the waveform that contains the junk radiation is discarded \cite{sxs25_103}.

\begin{figure}[!tbp]
    \centering
    \includegraphics[width=\columnwidth]{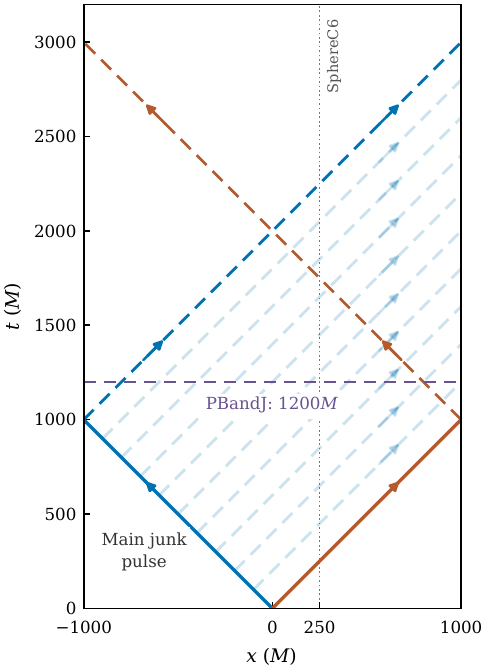}
    \caption{Schematic spacetime paths of the junk radiation, with position $x$ and time $t$ in units of the total binary mass $M$. Solid lines show the main junk-radiation fronts traveling outward from near the black holes, with blue and red indicating propagation in the negative and positive $x$ directions, respectively. The fronts travel at the speed of light and reach the outer boundaries at $t\approx 1000M$, where they are partially reflected back toward the origin (dashed lines of the same color). The semi-opaque dashed blue lines show reflections from subdomain boundaries between the SphereCs. The corresponding red reflections are omitted for clarity. These internal reflections stay for a long time in the computational domain, and we suspect that they cause the residual junk radiation. The purple horizontal line marks PBandJ at $t=1200M$. The dotted gray vertical line at $x=250M$ marks the approximate location of SphereC6. We expect the reflected junk-radiation front to pass through SphereC6 at $t\approx2250M$, after which the constraint violations should rapidly decrease.}
    \label{fig_junk_spacetime}
\end{figure}

Even after the main junk-radiation front has left a subdomain, we continue to see large constraint violations for much longer. We suspect that these large constraint violations are caused by the residual junk, which is produced by reflections of the main junk-radiation pulse from subdomain boundaries as it propagates through the computational domain, as illustrated in Figure~\ref{fig_junk_spacetime}. As the residual junk propagates away and the associated constraint violations are damped, the constraints gradually settle to a post-junk level determined primarily by the numerical resolution. The reflection from the outer boundary then propagates inward. For an outer boundary at approximately $1000M$, residual junk can remain in the domain for thousands of $M$. The red and blue dashed lines in Figure~\ref{fig_junk_spacetime} show the pulses reflected from the outer boundaries returning to the inner region around $t=2000M$ and continuing through the center toward the opposite outer boundary, which they reach around $t=3000M$.

Because of the junk radiation, the BBH parameters (masses, spins, and orbital parameters) we get from the initial data solver differ from those we measure after the initial junk transients have decayed. We call these the `relaxed' parameters. Since the junk radiation is unresolved, its behavior changes with resolution, so the relaxed parameters differ between levels even when we start from the same initial data. These slight differences affect the waveform mismatch and merger time, complicating comparisons between resolutions. To reduce this problem, we first evolve the system at the highest level we want until the main junk-radiation front has propagated out of the domain. We then branch this simulation into the lower levels and allow AMR to reduce the resolution accordingly. We call this process “perform branching after junk” (PBandJ) \cite{Scheel:2025jct}. For the simulations considered here, PBandJ was done at $t=1200M$, marked by the purple horizontal dashed line in Figure~\ref{fig_junk_spacetime}. Residual junk radiation is still present when we branch, but its amplitude is much smaller, so its effect on the relaxed parameters is small.

\subsection{Binary black hole system parameters}
\label{ssec:Binary black hole system parameters}

For this study, we evolved an equal-mass, non-spinning binary black hole (BBH) system through the inspiral. All times and distances are quoted in units of the total mass $M=M_A+M_B$, where $M_A$ and $M_B$ are the Christodoulou masses of the individual black holes. Inputs to the initial-data solver \cite{sxs25_141,sxs25_143} were
\begin{align*}
    M &= 1,\\
    q &= 1,\\
    d &= 18M,\\
    \Omega_0 &= 0.012226459838/M,\\
    \dot{a}_0 &= -1.72797015626\times10^{-5}/M,\\
    \chi_A^{\mathrm{input}} &= (0,0,0),\\
    \chi_B^{\mathrm{input}} &= (0,0,0).
\end{align*}

The resulting initial data solution had total mass $M_A+M_B=0.9999999426$ and ADM energy $E_{\mathrm{ADM}}=0.9934279614$. We show many digits as reported by SpEC for reproducibility, but the last few digits may be affected by numerical error. The individual masses and spins from the initial-data solver, together with the relaxed parameters measured after PBandJ, are listed in Appendix~\ref{app:Parameters of the systems}.

All simulations in this paper start from this same Superposed Kerr--Schild (SKS) initial data \cite{sxs25_134} with negative expansion boundary conditions \cite{sxs25_139}. We use damped harmonic gauge for all simulations \cite{sxs25_155,sxs25_156,sxs25_157}. No eccentricity reduction \cite{sxs25_144,sxs25_145,sxs25_146,sxs25_147,2026PhRvD.114d4071T} was performed and the system has a mild eccentricity of $\approx 0.05$. All levels were evolved from PBandJ at $t=1200M$ to $t=4000M$, and the black holes completed around $6.5$ orbits during this period.

\section{Accuracy metrics in SpEC}
\label{sec:Accuracy metrics in SpEC}

A BBH evolution can be roughly divided into three distinct phases: the inspiral, the merger, and the ringdown. Evolving each of these phases poses unique challenges and requires very different strategies to handle. In terms of coordinate time evolved per CPU hour, the merger and the ringdown parts of the simulation are slower than the inspiral part. But the inspiral part ends up consuming most computing resources because it runs the longest. We will focus on the inspiral phase of the simulation because most of the error accumulation happens during this phase \cite{Mitman_2025}.

In Section~\ref{ssec:Constraint violations and Constraint damping}, we discuss constraint violations and spectral convergence. In Section~\ref{ssec:Spectral coefficients}, we explain how spectral coefficients and power monitors can be used to check the accuracy of the fundamental variables $g_{ab}$ and $\kappa_{abc}$. Finally, in Section~\ref{ssec:Waveform Difference} we define waveform difference, the quantity we will use to estimate the errors in our waveforms.

\subsection{Constraint violations and constraint damping}
\label{ssec:Constraint violations and Constraint damping}

The first-order Generalized Harmonic (GH) formulation of Einstein's equations (Section~\ref{ssec:Generalized Harmonic System}) consists of evolution equations and constraint equations. In practice, we solve the constraint equations at the initial time, and we solve only the evolution equations to advance the solution to subsequent times, ignoring the constraints. This is allowed because of the following property: if the constraint equations are solved exactly at $t=0$, then solving the evolution equations exactly will guarantee that the constraint equations are satisfied at all subsequent times. In a numerical simulation, however, we never solve the evolution equations exactly, but only to some finite accuracy. So the constraint equations will not be satisfied exactly either. We check errors in our solution by monitoring the constraints. The full set of constraints are $\left\{\mathcal{C}_a, \mathcal{F}_a, \mathcal{C}_{i a}, \mathcal{C}_{i a b}, \mathcal{C}_{i j a b}\right\}$ in terms of the GH variables:
\begin{align}
    \mathcal{C}_a &= H_a+\Gamma_a \label{eq_constraint_1}, \\
    \mathcal{F}_a &\approx t^c \partial_c \mathcal{C}_a, \\
    \mathcal{C}_{i a} &\approx \partial_i \mathcal{C}_a, \\
    \mathcal{C}_{i a b} &= \partial_i g_{a b}-\Phi_{i a b} \label{eq_constraint_3}, \\
    \mathcal{C}_{i j a b} &= 2 \partial_{[i} \Phi_{j] a b}=2 \partial_{[j} \mathcal{C}_{i] a b}.
\end{align}
Here, $\approx$ means that we have omitted terms proportional to the constraints \cite{sxs25_153}.

The individual constraints have different physical dimensions, so mathematically we need appropriate factors of the total mass $M$ before adding them. In SpEC, the total mass in code units remains close enough to one during the evolution that we omit these factors. While looking at the individual constraints is informative, it is helpful to define a generalized-harmonic constraint energy density, $\mathcal{E}_{C}$ that captures the effect of all these individual constraints:
\begin{equation}
\begin{aligned}
\mathcal{E}_{C}
&= \bigg[
 \delta^{ab}\mathcal{C}_a \mathcal{C}_b
 + \delta^{ab}\mathcal{F}_a \mathcal{F}_b
 + \gamma^{ij}\delta^{ab}\mathcal{C}_{i a} \mathcal{C}_{j b} \\
&\quad + \gamma^{ij}\delta^{ac}\delta^{bd}\mathcal{C}_{i a b} \mathcal{C}_{j c d}
 + \gamma^{ij}\gamma^{kl}\delta^{ac}\delta^{bd}\mathcal{C}_{i k a b} \mathcal{C}_{j l c d}
 \bigg]^{1/2},
\end{aligned}
\label{eq_GhCe}
\end{equation}
where $\gamma^{ij}$ is the inverse spatial metric. Whenever we talk about constraint violations without qualifications, we mean the quantity $\mathcal{E}_{C}$.

Some fundamental variables in the GH formalism, like $\kappa_{011}$, go from being of order unity near the black holes to being of order $10^{-6}$ at the outer boundary. This makes $\mathcal{E}_{C}$ unsuitable for comparing errors in different parts of the computational domain with each other. We therefore define a normalization, $\mathcal{N}_{C}$, as
\begin{equation}
\begin{aligned}
    \mathcal{N}_{C}
    &= \bigg[
    \gamma^{ij}\delta^{ac}\delta^{bd} \partial_{i} g_{ab} \partial_{j} g_{cd} \\
    &\quad + \gamma^{ij}\delta^{ac}\delta^{bd} \partial_{i} \Pi_{ab} \partial_{j} \Pi_{cd} \\
    &\quad + \gamma^{ij}\gamma^{kl}\delta^{ac}\delta^{bd} \partial_{i} \Phi_{kab} \partial_{j} \Phi_{lcd}
    \bigg]^{1/2}.
\end{aligned}
\label{eq_GhCe_normalization}
\end{equation}
Here we have again omitted factors of $M\approx1$. Using this normalization, we can define the normalized constraint energy density, $\widehat{\mathcal{E}}_{C}$, as
\begin{equation}
\begin{aligned}
     \widehat{\mathcal{E}}_{C} = \frac{\mathcal{E}_{C}}{\mathcal{N}_{C}}.
\end{aligned}
\label{eq_NormalizedGhCe}
\end{equation}

Because of the finite resolution used in numerical simulations, the constraint violations will always have a small non-zero value. Such small constraint violations can grow rapidly as we evolve the system forward in time, causing the simulation to fail. Constraint damping prevents this catastrophic growth by damping the small constraint violations and is essential for a stable evolution \cite{sxs25_153}. In the Generalized Harmonic formalism (Appendix~\ref{app:Generalized Harmonic System}), constraint damping requires a positive value of $\gamma_0$ and $\gamma_2$. $\gamma_0$ controls the time scale over which the 1-index constraints $C_a$ are exponentially damped and $\gamma_2$ does the same for the 3-index constraints $C_{iab}$. The other constraint violations are related to derivatives of these two constraints, and choosing positive $\gamma_0$ and $\gamma_2$ also helps suppress them \cite{sxs25_153}.

While one can choose the values of $\gamma_0$ and $\gamma_2$ separately, in SpEC they are usually chosen to be the same for simplicity. The maximum growth in the constraints happens near the black holes, where the fundamental variables are changing rapidly. Therefore, compared to the outer subdomains, the values of $\gamma_0$ and $\gamma_2$ near the black holes are chosen to be much larger with a Gaussian falloff. Let $m_A$ and $m_B$ denote the individual black-hole Christodoulou masses, $M=m_A+m_B$ the total mass, and $d$ the initial coordinate separation. Using the coordinate distances defined in Eq.~\eqref{eq:coordinate_distances}, the spatial profile is
\begin{align}
\gamma_0(\mathbf{x})
= \gamma_2(\mathbf{x})
&=
\frac{4}{m_A} \exp\!\left[-\frac{r_A^2}{\left(7m_A\right)^2}\right]
\nonumber \\
&+
\frac{4}{m_B} \exp\!\left[-\frac{r_B^2}{\left(7m_B\right)^2}\right]
\nonumber \\
&+
\frac{0.075}{M} \exp\!\left[-\frac{r^2}{\left(2.5d\right)^2}\right]
\nonumber \\
&+
\frac{0.001}{M}.
\label{eqn:constraint damping gamma values spatial profile}
\end{align}
For the equal-mass BBH system considered here, $m_A\simeq m_B\simeq M/2$ and $M\simeq1$. For the initial separation $d=18M$, the two near-hole Gaussians in Eq.~\eqref{eqn:constraint damping gamma values spatial profile} have amplitudes of approximately $8$ and widths of approximately $3.5$. The broader Gaussian has amplitude $0.075$ and width $45$, and the constant damping floor is $0.001$.

\subsection{Spectral coefficients and power monitors}
\label{ssec:Spectral coefficients}
One advantage of using spectral methods is that we have access to the spectral coefficients and power monitors of our fundamental GH variables $g_{ab}$ and $\kappa_{abc}$. For a smooth, well-resolved function, successive power monitors in any spectral basis should decrease exponentially with mode number. For example, for the radial spectral basis defined above, $P^{(r)}_k\propto e^{-\alpha k}$ for some $\alpha>0$. The truncation error defined in Eq.~\eqref{eqn:Truncation Error} uses these power monitors to estimate how well a particular variable is resolved in a subdomain. This gives us more fine-grained information than the constraint violations. In particular, constraint violations indicate that quantities that should agree are inconsistent, but they do not tell us which quantity is inaccurate.

\subsection{Waveform Difference}
\label{ssec:Waveform Difference}
Another measure of accuracy that we will focus on is the waveform difference $\Delta$ \cite{Scheel:2025jct}. This quantity measures the difference between two waveforms over a chosen time interval and angular sphere and is defined as
\begin{equation}
\begin{aligned}
& \Delta\left(h_1, h_2\right)=\frac{1}{2} \frac{\left\|h_1-h_2\right\|^2}{\sqrt{\left\|h_1\right\|^2\left\|h_2\right\|^2}} \\
& \left\|h_1\right\|=\sqrt{\int_{t_1}^{t_2} \int_{S^2}\left|h_1\right|^2 d \Omega d t}
\end{aligned}
\label{eq_waveform_difference}
\end{equation}
Here $h_1$ and $h_2$ are the gravitational waveforms from two simulations, in our case, two different levels. For either waveform, orthonormality of the spherical harmonics gives
\begin{equation}
    \int_{S^2}|h|^2\,d\Omega=\sum_{\ell,m}|h_{\ell m}(t)|^2,
    \label{eq_waveform_angular_integral}
\end{equation}
where $h_{\ell m}$ are the coefficients of the expansion of $h$ in spin-weight $-2$ spherical harmonics ${}_{-2}Y_{\ell m}$. We extract all modes with $2\leq\ell\leq8$ and $-\ell\leq m\leq\ell$, and include them all in the waveform difference unless stated otherwise. For a flat noise power spectral density and in the limit of infinite numerical resolution, the waveform difference approaches the mismatch averaged over the two-sphere \cite{Scheel:2025jct}. We use waveform difference to avoid detector-dependent frequency weighting, which can emphasize errors in different parts of the waveform \cite{2026PhRvD.113l4001K}.

Even when evolving the same initial data at different resolutions, we do not expect to have a zero waveform difference because of various numerical errors. However, just like constraint violations, we expect the difference to improve as we increase the resolution.

For waveform differences, we use gravitational waveforms extracted at a radius of $257M$ using the Regge--Wheeler--Zerilli extraction procedure \cite{sxs25_96,sxs25_97,sxs25_98,sxs25_99,sxs25_100}. The gravitational waveforms in the SXS catalog \cite{Scheel:2025jct}, however, are computed by extrapolating a series of these finite-radius Regge--Wheeler--Zerilli waveforms to future null infinity and then correcting for center-of-mass effects and gravitational-wave memory effects. Because the goal of this paper is to study the errors in the methods of solving the Einstein equations and not the errors in processing and correcting the waveforms, we choose to analyze the simpler finite-radius waveforms here. For the same reason, we do not align the waveforms or perform any optimization when computing the waveform difference. Keep in mind, however, that for physics applications one should use the waveforms computed by the extrapolation/correction procedure, or even better by Cauchy-Characteristic Extraction (CCE) \cite{sxs25_176,sxs25_177,sxs25_178,sxs25_179,sxs25_180,sxs25_181}.

\begin{figure*}
    \centering
    \captionsetup[subfloat]{position=top,font=normalsize,justification=centering,singlelinecheck=true,skip=6pt}
    \subfloat[\label{fig_L15_main_L2(GhCe)} Old: full domain]{
        \includegraphics[width=0.47\linewidth]{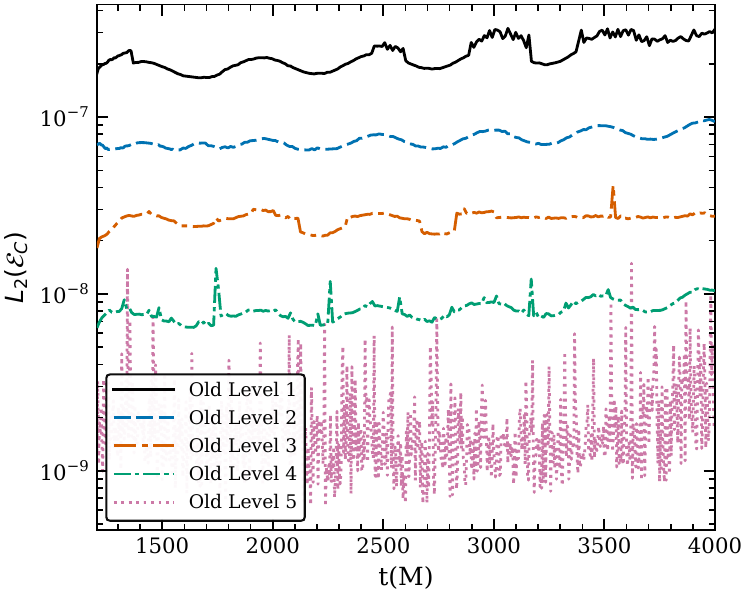}
    }
    \hfill
    \subfloat[\label{fig_L15_main_SphereA0_Linf_GhCe} Old: SphereA0]{
        \includegraphics[width=0.47\linewidth]{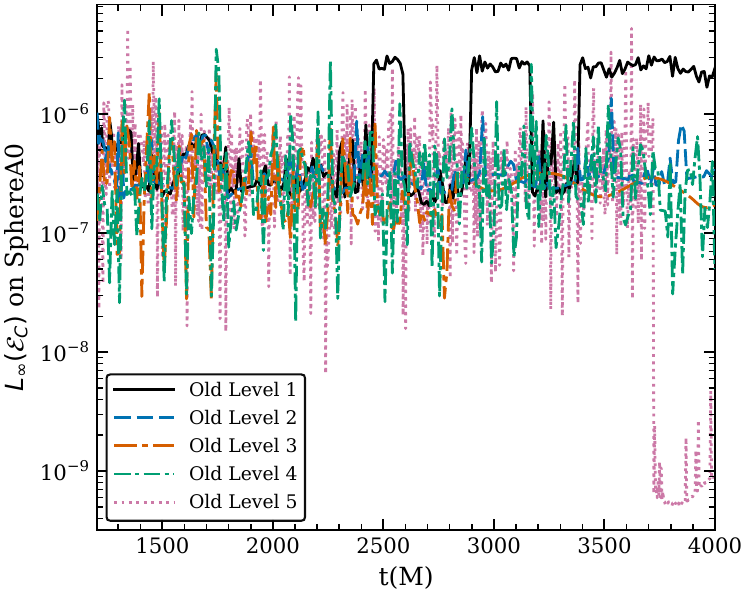}
    }
    \\[0.7em]
     \subfloat[\label{fig_L15_ode_fix_L2(GhCe).pdf} ODE fix: full domain]{
         \includegraphics[width=0.47\linewidth]{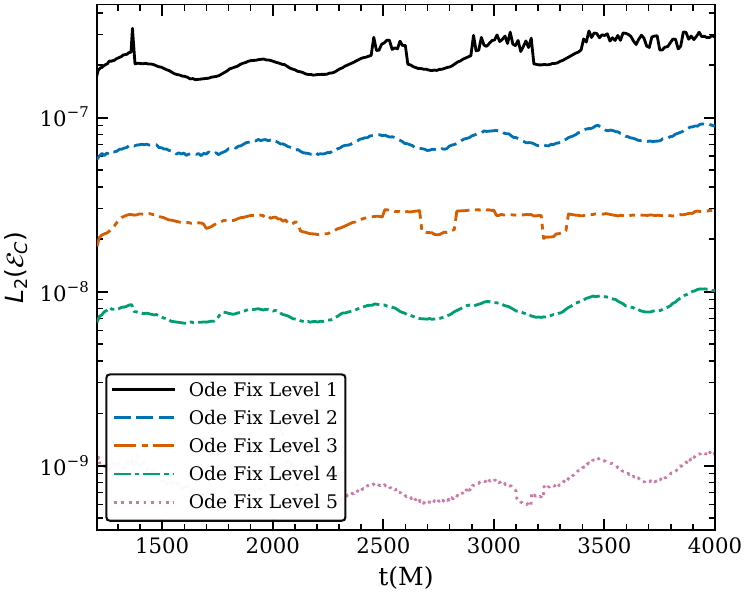}
    }
    \hfill
    \subfloat[\label{fig_L15_ode_fix_SphereA0_Linf_GhCe.pdf} ODE fix: SphereA0]{
        \includegraphics[width=0.47\linewidth]{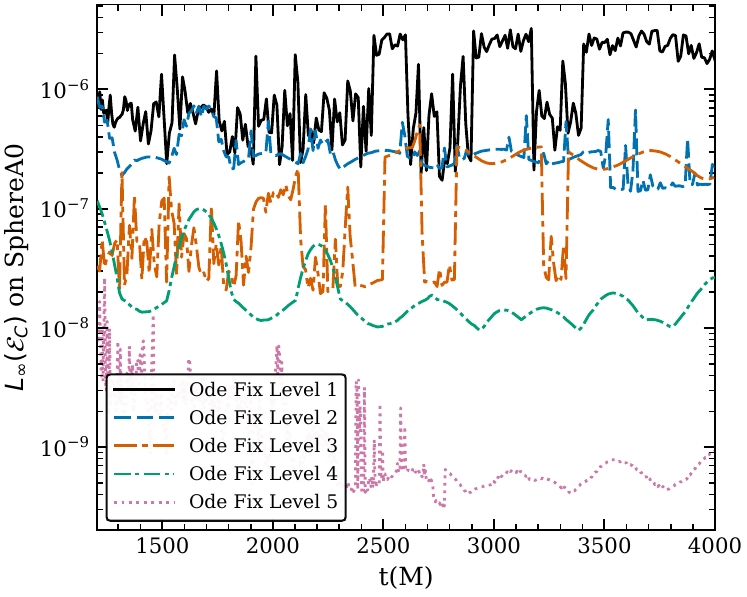}
    }
    \caption{\label{fig_near_horizon_constraints_ode_fix} Constraint violations before and after changing the ODE tolerance for levels 1--5. The top row shows the Old simulations, and the bottom row shows the simulations with the ODE tolerance fix. The left panels show the $L_2$ norm over the full computational domain, while the right panels show the $L_\infty$ norm in SphereA0. The Old simulations show noisy high-resolution behavior near the horizon, while the simulations with the ODE tolerance fix show cleaner convergence.}
\end{figure*}

\section{Improvements to SpEC}
\label{sec:Improvements to SpEC}
In this section, we describe the changes we made to SpEC to improve the accuracy metrics discussed in Section~\ref{sec:Accuracy metrics in SpEC}. In Section~\ref{ssec:Adaptive ODE Tolerance for Near-Horizon Constraint Convergence}, we describe how changing the ODE tolerance adaptively improved convergence near the black hole horizons. In Section~\ref{ssec:More Constraint Damping and Resolution in the Outer Subdomain}, we discuss how more constraint damping and grid resolution improved the constraint violations in the outer subdomains. In Section~\ref{ssec:More Resolution in the Connecting Subdomains for better Waveform Difference}, we show that finite-radius waveform differences improve by around two orders of magnitude when we add more resolution to the subdomains connecting the two black holes. We also show that the waveform differences are strongly correlated with deviations in the black holes' trajectories across different resolutions. Finally, in Section~\ref{ssec:Computational Cost of Improvements} we discuss the computational cost of these improvements and explain why the cost increase is modest compared to the accuracy gain.

In the following sections, we compare three sets of simulations. We use the following labels:
\begin{itemize}
    \item ``Old'' refers to simulations run with the main branch of SpEC, using the spatially varying AMR tolerance in Eq.~\eqref{eq:spatial_amr_tolerance}.
    \item ``ODE fix'' refers to simulations in which the ODE tolerance is changed from the default value of $10^{-8}$ to $\epsilon_{\mathrm{base}}/2000$. This makes the ODE tolerance adaptive, allowing it to change with the AMR tolerance.
    \item ``New'' refers to simulations that use the adaptive ODE tolerance, increase the asymptotic value of the constraint damping from $0.001$ to $0.01$, and use the spatially uniform AMR tolerance defined in Eq.~\eqref{eq:uniform_amr_tolerance}.
\end{itemize}

\begin{figure}
    \includegraphics[width=\linewidth]{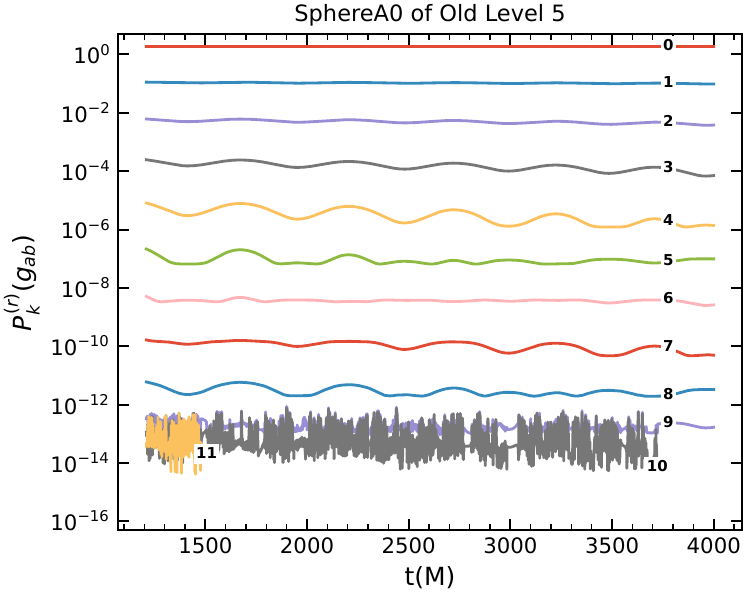}
    \caption{\label{fig_L15_main_L5_PS_psi_SphereA0_0} Radial power monitors $P^{(r)}_k(g_{ab})$ versus time in SphereA0 for an Old level 5 simulation. The amplitudes decrease approximately exponentially with $k$ through the lower modes, while the highest modes are dominated by noise.}
\end{figure}

\begin{figure}
    \includegraphics[width=\linewidth]{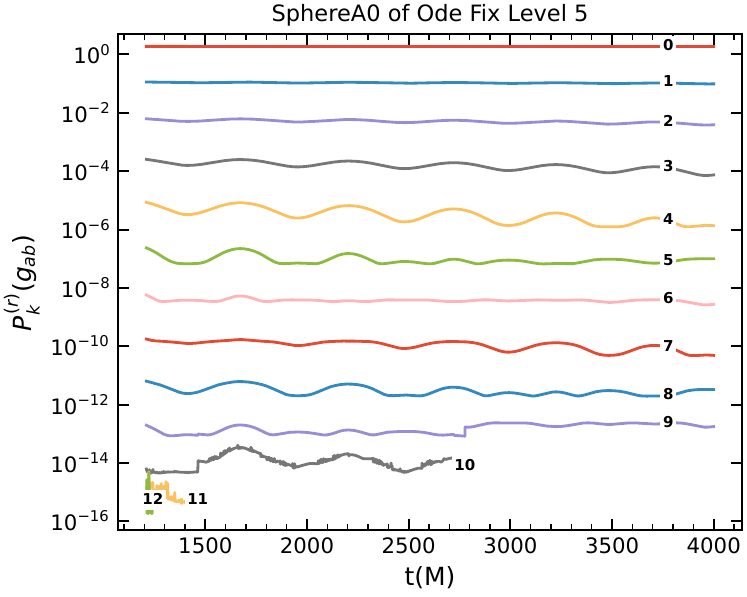}
    \caption{\label{fig_L15_ode_fix_L5_PS_psi_SphereA0_0} Radial power monitors $P^{(r)}_k(g_{ab})$ versus time in SphereA0 for an ODE-fix level 5 simulation. Compared with the Old simulation, the highest retained modes contain less noise, and the AMR algorithm drops noisy modes.}
\end{figure}

\subsection{Adaptive ODE Tolerance for Near-Horizon Constraint Convergence}
\label{ssec:Adaptive ODE Tolerance for Near-Horizon Constraint Convergence}

When discussing constraint violations, we focus on SphereA0 near the black holes and SphereC6 in the wave extraction region. These subdomains are shown in Figure~\ref{fig_SpEC_subdomains} and defined in Section~\ref{ssec:Computational domain in SpEC}.

Figure~\ref{fig_L15_main_L2(GhCe)} shows the $L_2$ norm of $\mathcal{E}_{C}$  over the whole computational domain for different levels. We can see that the $L_2$ norm goes down as we increase the resolution. While the simulations are converging, there is a lot of noise in the highest-resolution level 5 simulation. To better understand the source of this noise, we need to look at each subdomain separately. We will primarily look at the $L_\infty$ norm of the constraints because the $L_2$ norm averages over all the points, hiding issues we might have in some local region of the computational domain.

For these simulations, the noise is present in the subdomains SphereA0 (Figure~\ref{fig_L15_main_SphereA0_Linf_GhCe}) and SphereB0 (not shown). The constraint violations in SphereA0 do not converge at all, particularly as we go to higher resolutions like level 4 and level 5. After some investigation, we discovered that this non-convergence is caused by SpEC not scaling the fixed ODE solver tolerance correctly for high-resolution runs. By using the adaptive ODE tolerance, we can see a clean exponential convergence of the constraint violations in SphereA0 (Figure~\ref{fig_L15_ode_fix_SphereA0_Linf_GhCe.pdf}). Furthermore, by comparing Figures~\ref{fig_L15_main_L2(GhCe)} and~\ref{fig_L15_ode_fix_L2(GhCe).pdf}, we can see that the noise in the $L_2$ norm of $\mathcal{E}_{C}$ over the whole domain is reduced as well, most notably in the high-resolution simulations.

The adaptive ODE tolerance formula gives a tolerance smaller than the previous value of $10^{-8}$ for levels 2 and above. Since the noise in the constraints decreased as we made the ODE tolerance smaller, we think that it was caused by the ODE stepper occasionally taking steps that violated the CFL limit. There is a sweet spot for the ODE tolerance where this noise is eliminated, but the average step size hardly decreases. In this regime, we get more accurate simulations without any extra computational cost.

\begin{figure}
    \includegraphics[width=\linewidth]{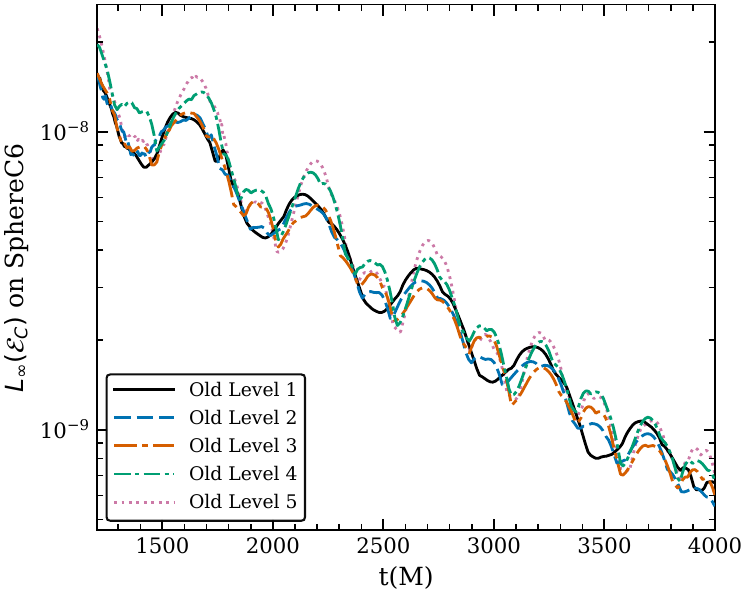}
    \caption{\label{fig_L15_main_SphereC6_Linf_GhCe} The $L_\infty$ norm of the constraint energy density, $L_\infty(\mathcal{E}_{C})$, in SphereC6 versus time for Old simulations at levels 1--5.}
\end{figure}

\begin{figure*}
    \centering
    \captionsetup[subfloat]{position=top,font=normalsize,justification=centering,singlelinecheck=true,skip=6pt}
    \subfloat[\label{fig_L15_main_SphereC6_Linf_NormalizedGhCe.pdf} Old]{
        \includegraphics[width=0.47\linewidth]{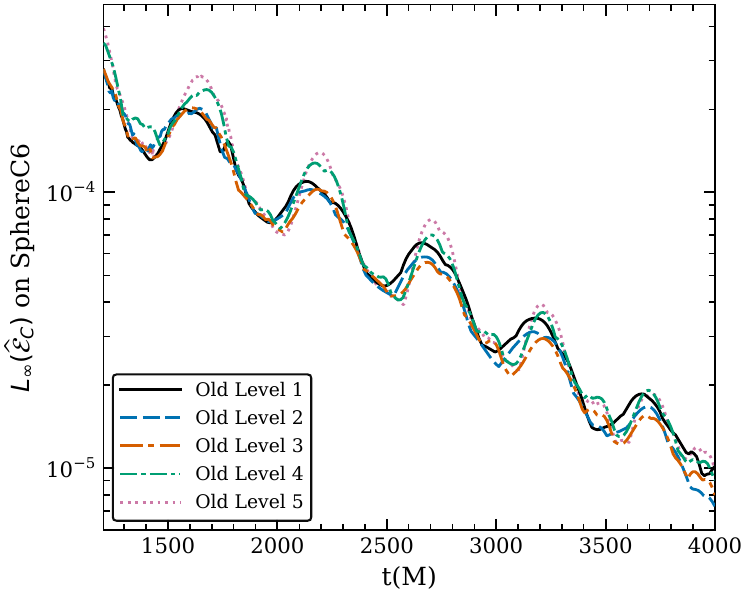}
    }
    \hfill
    \subfloat[\label{fig_L15_ode_fix_SphereC6_Linf_NormalizedGhCe.pdf} ODE fix]{
        \includegraphics[width=0.47\linewidth]{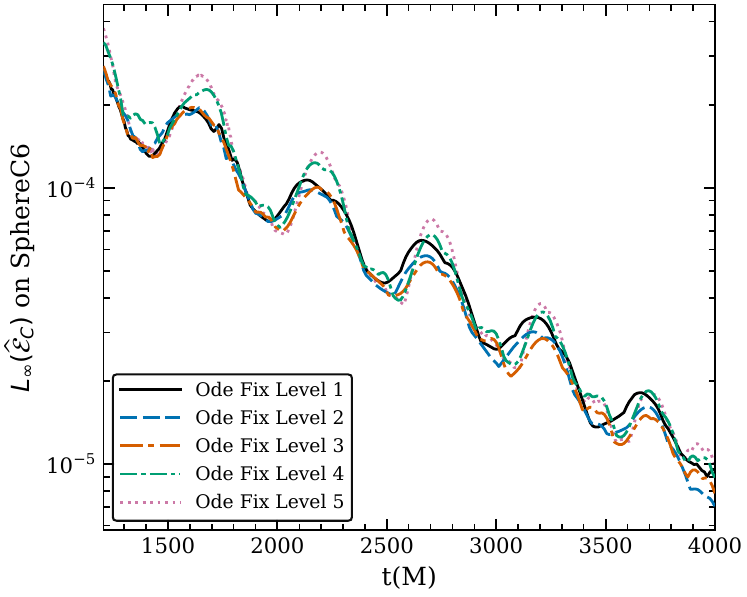}
    }
    \caption{\label{fig_spherec6_normalized_constraints_ode_fix} The $L_\infty$ norm of the normalized constraint energy density, $L_\infty(\widehat{\mathcal{E}}_{C})$, in SphereC6 for levels 1--5 before and after the ODE tolerance fix. The ODE tolerance fix improves the near-horizon convergence, but the normalized constraints in the wave-extraction region still show poor convergence with level. This motivates adding both more resolution and more constraint damping in the outer subdomains.}
\end{figure*}

\begin{figure}
    \includegraphics[width=\linewidth]{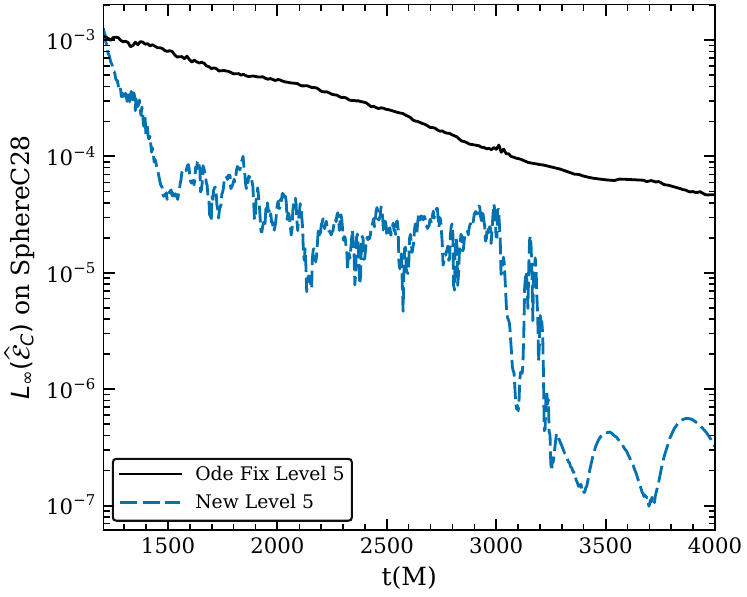}
    \caption{\label{joined_ML_5_S1_L5_SphereC28_Linf_NormalizedGhCe.pdf} The $L_\infty$ norm of the normalized constraint energy density, $L_\infty(\widehat{\mathcal{E}}_{C})$, in SphereC28 ($r\approx950M$) versus time for the ODE-fix and New level 5 simulations. The ODE-fix result continues to decay throughout the plotted interval, whereas the New result saturates earlier because of its stronger constraint damping.}
\end{figure}

To find the optimal ODE tolerance for the simulations, we decreased the tolerance and chose the largest value that prevented CFL violations without reducing the average step size. We found that this value is roughly proportional to the AMR tolerance. This led us to the heuristic prescription $\epsilon_{\mathrm{ODE}}=\epsilon_{\mathrm{base}}/2000$. This formula may not be optimal for more complex simulations, which may require more grid points to reach the same AMR tolerance. Also, during very long simulations, the domain changes substantially as the black holes get closer, so the optimal ODE tolerance can change with time. We are therefore looking into more dynamic strategies that automatically select the ODE tolerance based on the current simulation grid.

While the improvements in the constraint violations are encouraging, it is also instructive to compare the power monitors before and after the ODE tolerance change. Consider Figure~\ref{fig_L15_main_L5_PS_psi_SphereA0_0}, which shows the power spectrum of $g_{ab}$ in the SphereA0 subdomain for a level 5 run. We can immediately see that the two highest radial power monitors, $P^{(r)}_{10}(g_{ab})$ and $P^{(r)}_{11}(g_{ab})$, are dominated by noise, which in part explains the large constraint violations in Figure~\ref{fig_L15_main_SphereA0_Linf_GhCe}. Comparing this with Figure~\ref{fig_L15_ode_fix_L5_PS_psi_SphereA0_0}, which includes the ODE tolerance fix, we see that $P^{(r)}_{10}(g_{ab})$ has less noise and is dropped by the AMR algorithm earlier, around $2700M$. The noise in $P^{(r)}_{10}(g_{ab})$ in Figure~\ref{fig_L15_main_L5_PS_psi_SphereA0_0} is several orders of magnitude smaller than the constraint violations in Figure~\ref{fig_L15_main_SphereA0_Linf_GhCe}. This is because constraint violations involve derivatives of the fundamental variables, which amplify the error.

Looking at Figure~\ref{fig_L15_ode_fix_L5_PS_psi_SphereA0_0}, we see that $P^{(r)}_{0}(g_{ab})\approx1$ and $P^{(r)}_{9}(g_{ab})\approx10^{-12}$. This means that in SphereA0 we are resolving contributions to $g_{ab}$ at roughly the twelfth decimal place. Later, we will look at level 6 simulations where even the $P^{(r)}_{10}(g_{ab})$ mode is resolved (Figure~\ref{fig_L16_set1_L6_PS_psi_SphereA0_0}), taking us to the thirteenth decimal place and beyond. It is natural to stop and ask why we need so many significant digits in the metric to achieve the waveform accuracy required by next-generation GW detectors. Part of the answer lies in the fact that extracting gravitational waves requires derivatives of the metric, which amplify numerical errors. We also have to consider that, for a typical SpEC simulation, the ODE solver takes approximately $10^6$ time steps, introducing numerical errors that can accumulate over time. There are also other sources of error, such as the AMR algorithm and horizon tracking. Therefore, even with more than ten significant digits, we may not have the headroom one might assume just by looking at the waveform accuracy requirement. In fact, from the simulations in the SXS catalog \cite{Scheel:2025jct}, we know that there are systems where the metric components in the inner subdomains are resolved to the tenth significant digit, but the final waveform mismatches are still too large for next-generation GW detectors.

\begin{figure}
    \includegraphics[width=\linewidth]{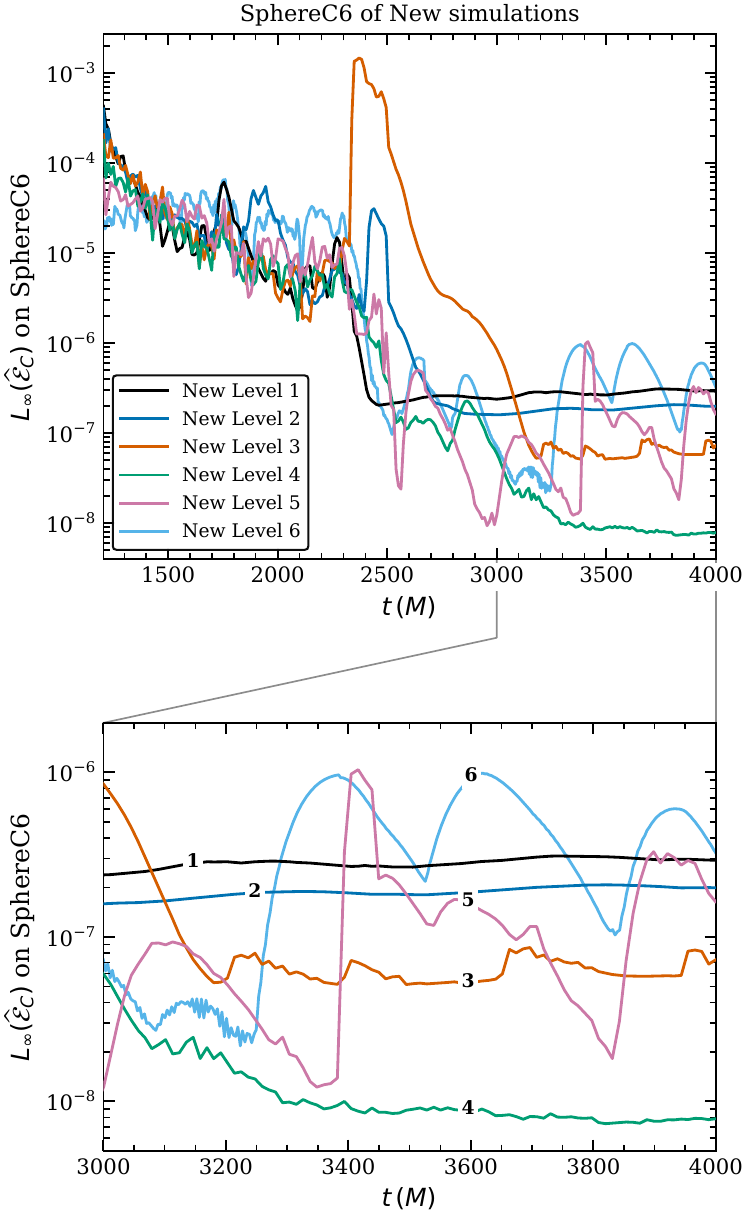}
    \caption{\label{fig_L16_set1_SphereC6_Linf_NormalizedGhCe} The $L_\infty$ norm of the normalized constraint energy density, $L_\infty(\widehat{\mathcal{E}}_{C})$, in SphereC6 versus time for New simulations at levels 1--6. The lower panel shows a zoom of $3000M\leq t\leq4000M$, with numbers identifying the resolution levels.}
\end{figure}

\begin{figure}
    \includegraphics[width=\linewidth]{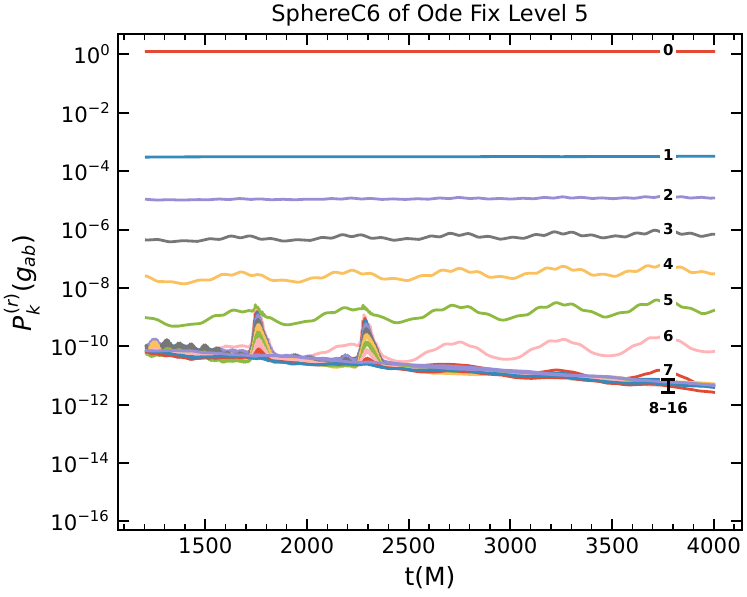}
    \caption{\label{fig_L15_ode_fix_L5_PS_psi_SphereC6_0} Radial power monitors $P^{(r)}_k(g_{ab})$ versus time in SphereC6 for an ODE-fix level 5 simulation.}
\end{figure}

\begin{figure}
    \includegraphics[width=\linewidth]{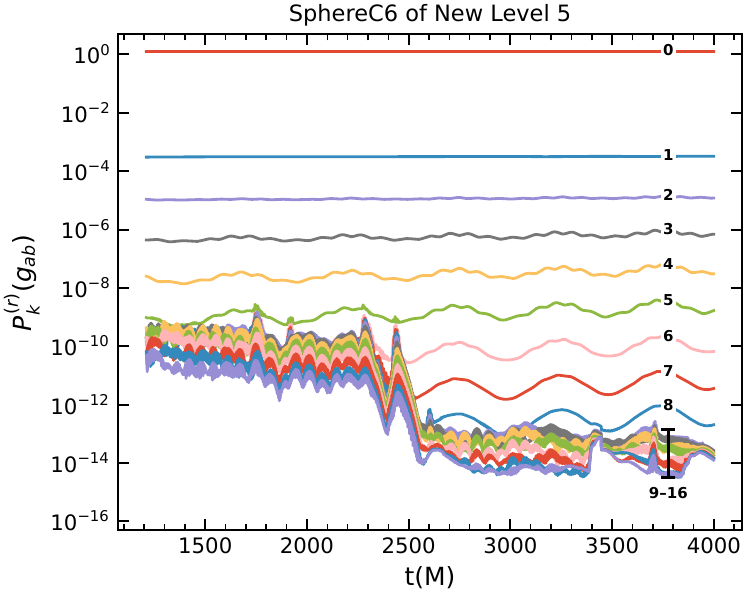}
    \caption{\label{fig_L16_set1_L5_PS_psi_SphereC6_0} Radial power monitors $P^{(r)}_k(g_{ab})$ versus time in SphereC6 for a New level 5 simulation.}
\end{figure}

\subsection{More Constraint Damping and Resolution in the Outer Subdomain}
\label{ssec:More Constraint Damping and Resolution in the Outer Subdomain}

Having resolved the issue in the subdomains near the black holes, we can now focus on the subdomains in the wave extraction region. For now, we will focus on the subdomain SphereC6, one of the first subdomains from which we extract the gravitational wave data. The constraint violations in SphereC6 are shown in Figure~\ref{fig_L15_main_SphereC6_Linf_GhCe}. Note that these constraint violations are small compared to the constraint violations in the inner subdomain, although there is hardly any convergence as we increase the level. Furthermore, we cannot compare the constraint violations in two different subdomains without normalization because the fields may have very different values in those subdomains. To account for this, we will look at the normalized constraint energy density $\widehat{\mathcal{E}}_{C}$ defined in Eq.~\eqref{eq_NormalizedGhCe}. The normalized constraint violations for SphereC6 are shown in Figure~\ref{fig_spherec6_normalized_constraints_ode_fix}. Compared to the unnormalized constraint violations, the normalized constraint violations are smaller by a factor of $10$ in the inner subdomains and larger by a factor of $10^4$ in the regions we extract our gravitational waves from. This is because the fundamental fields fall off rapidly as we move away from the location of the black holes, and normalizing the constraints captures this effect.

As explained in Section~\ref{ssec:AMR tolerance and definition of levels}, we try to have more resolution in the inner subdomains because it is the most nonlinear region of the simulation and the region where gradients of the fundamental variables are the largest. However, the large normalized constraint violations in the outer regions suggest that we should have more resolution in the outer subdomains. We did a few test simulations and found that simply increasing the resolution in the outer subdomains does not help until very late into the simulations. From the solid black ODE-fix level 5 curve for SphereC28 in Figure~\ref{joined_ML_5_S1_L5_SphereC28_Linf_NormalizedGhCe.pdf}, we can see that the constraints are going down with time, but the rate is very slow. This tells us that we need both more resolution and more constraint damping in the outer subdomains.

The amount of constraint damping also changes spatially, as shown in Eq.~\eqref{eqn:constraint damping gamma values spatial profile}. The inner subdomains have more constraint damping than the outer subdomains. Because the asymptotic damping value is small, the constraints in the outer subdomains decay slowly. In short simulations, the run can end before the noise in the outer subdomains has fully decayed. To deal with this, we increased the asymptotic value of the constraint damping in Eq.~\eqref{eqn:constraint damping gamma values spatial profile} from $0.001$ to $0.01$. With this extra constraint damping and the increased resolution, we were able to reduce the constraint violations in the outer subdomains, as shown in Figure~\ref{fig_L16_set1_SphereC6_Linf_NormalizedGhCe}. Note, however, that the constraint violations for levels 5 and 6 are worse than those for level 4. We strongly suspect this is caused by noise in the AMR procedure, described in Appendix~\ref{app:Glitches due to grid changes}, because it occurs irregularly and only in some subdomains.

A natural question is why we cannot increase the asymptotic value of the constraint damping even further. There are two reasons. First, we find that stronger constraint damping makes the junk radiation worse as it propagates through the outer subdomains. Second, stronger constraint damping can make the constraints worse in some situations, as we will see in Section~\ref{ssec:Difference in scales of the evolution variables}.

It is also instructive to look at the power-monitor plots in Figures~\ref{fig_L15_ode_fix_L5_PS_psi_SphereC6_0} and~\ref{fig_L16_set1_L5_PS_psi_SphereC6_0}. We can see that after $t\approx2500M$, the modes $P^{(r)}_{7}(g_{ab})$ and $P^{(r)}_{8}(g_{ab})$ in Figure~\ref{fig_L16_set1_L5_PS_psi_SphereC6_0} are much better resolved, while those modes in Figure~\ref{fig_L15_ode_fix_L5_PS_psi_SphereC6_0} are still decaying slowly. Note that even in the New case, we reach the noise floor by radial mode $k=9$, but the AMR algorithm still retains modes up to $k=16$. This is due to a historical decision made during the early development of SpEC, which we are reconsidering for future simulations.

\subsection{More Resolution in the Connecting Subdomains for better Waveform Difference}
\label{ssec:More Resolution in the Connecting Subdomains for better Waveform Difference}

Consider the finite-radius waveform difference between different levels during the early inspiral, starting at coordinate time $1200M$ (i.e., after most of the junk radiation has passed) and ending at $4000M$. We are looking at the waveforms just after PBandJ, so the phase errors have not had enough time to accumulate and increase the waveform difference \cite{Mitman_2025}. This means that the waveform differences we see are not typical of those from a full simulation, which usually has many more orbits for such parameters; however, they are still useful for diagnosing improvements in the algorithm.

Table~\ref{tab_waveform_difference} shows the waveform difference for the Old and New runs. We can see that the waveform differences have improved by around two orders of magnitude.

\begin{figure}[!htbp]
    \centering
    \includegraphics[width=\linewidth]{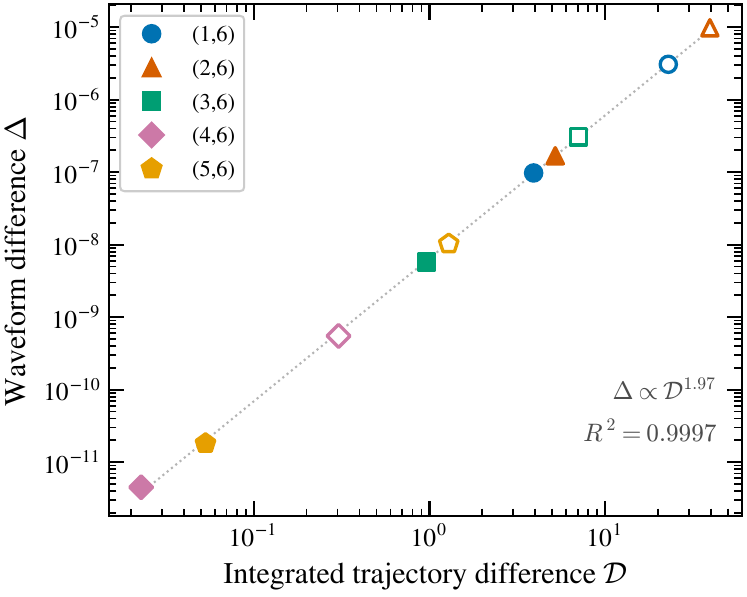}
    \caption{\label{fig_waveform_difference_combined} Waveform difference $\Delta$, defined in Eq.~\eqref{eq_waveform_difference}, versus the integrated trajectory difference $\mathcal{D}$, defined in Eq.~\eqref{eq:integrated_trajectory_difference}, for the Old and New simulations. Marker shapes and colors identify the level pairs listed in the legend. Filled markers show the New simulations and hollow markers show the Old simulations. We can see that all the points lie nearly on the same line, and each filled marker has a lower waveform difference than the corresponding hollow marker. The faint dotted line is a least-squares fit to all ten points in log space, $\log_{10}\Delta=a+b\log_{10}\mathcal{D}$, with $b=1.97$ and $R^2=0.9997$. This shows a very strong correlation between $\mathcal{D}$ and $\Delta$, with $\Delta$ scaling approximately as $\mathcal{D}^2$ during the inspiral part studied here.}
\end{figure}

\begin{figure*}[!t]
    \centering
    \captionsetup[subfloat]{position=top,font=normalsize,justification=centering,singlelinecheck=true,skip=6pt}
    \subfloat[\label{fig_waveform_difference_main} Old]{
        \includegraphics[width=0.47\linewidth]{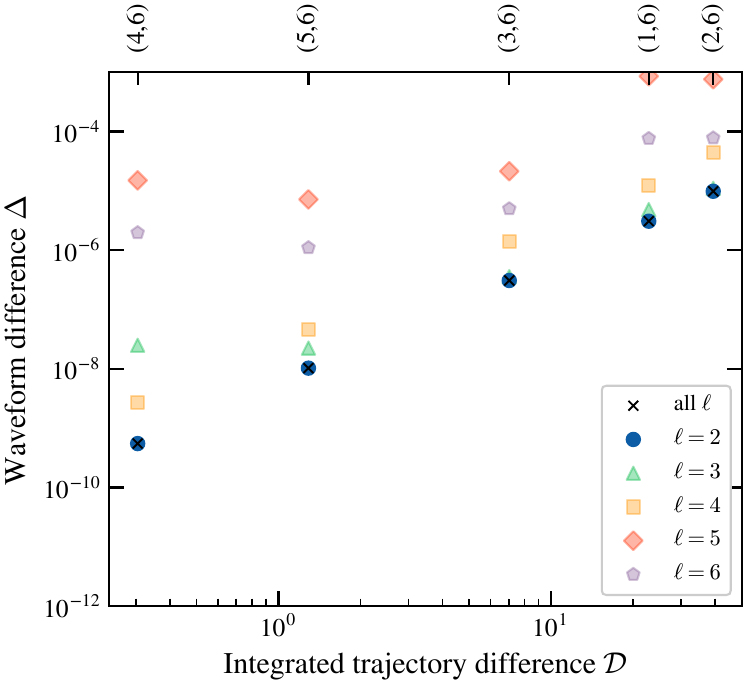}
    }\hfill
    \subfloat[\label{fig_waveform_difference_set1} New]{
        \includegraphics[width=0.47\linewidth]{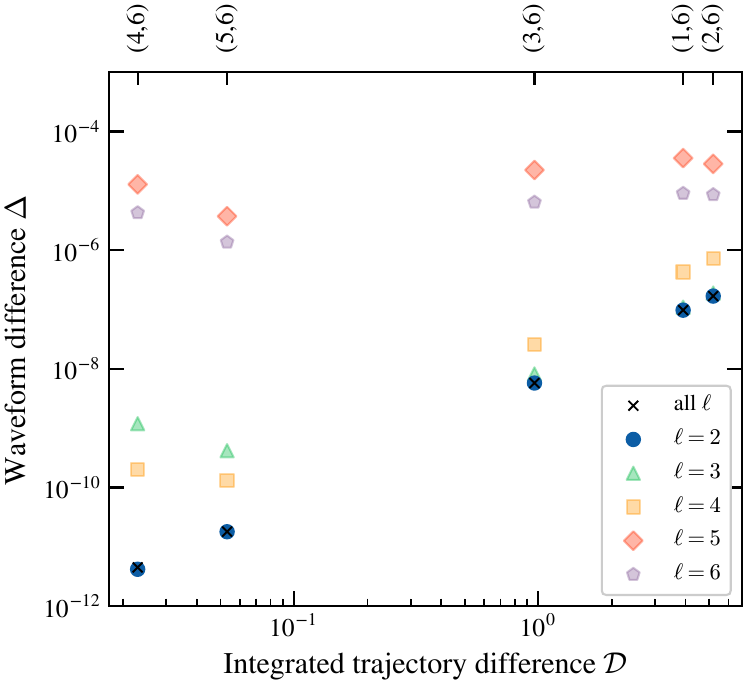}
    }
    \caption{\label{fig_waveform_difference_trajectory_correlation} Correlation between finite-radius waveform differences and differences in the black-hole trajectories. The bottom horizontal axis shows the time-integrated trajectory difference $\mathcal{D}^{(i,j)}$, defined in Eq.~\eqref{eq:integrated_trajectory_difference}, while the top horizontal axis identifies the corresponding pair of levels $(i,j)$. The vertical axis shows the finite-radius waveform difference $\Delta$, defined in Eq.~\eqref{eq_waveform_difference}. For each series labeled $\ell$, the sum in Eq.~\eqref{eq_waveform_angular_integral} includes only modes with that $\ell$ and all $m=-\ell,\ldots,\ell$, in both the numerator and denominator of $\Delta$. The waveform difference computed using only $\ell=2$ is visually indistinguishable from that computed using all extracted modes because the $(2,2)$ mode dominates. The $\ell=7$ and $\ell=8$ results are omitted for clarity, as they show no improvement between the Old and New simulations and behave similarly to the $\ell=5$ and $\ell=6$ results. In both the Old and New simulations, larger trajectory differences correspond to larger waveform differences. The New simulations reduce the waveform differences by approximately two orders of magnitude.}
\end{figure*}

Because waveform difference is used as the main predictor of the accuracy of a simulation for data analysis, it is crucial to understand how to reduce it. The general expectation is that the waveform difference should decrease with increasing resolution. We can see that this is mostly true in Table~\ref{tab_waveform_difference}. But note that in both cases, the waveform difference between level 2 and level 6 is higher than the waveform difference between level 1 and level 6, even though the constraint violations improve as we increase the level. The same behavior occurs when comparing levels 4 and 5 against level 6: the waveform difference between levels 4 and 6 is smaller than that between levels 5 and 6. This behavior is because the waveform difference is more strongly correlated with the difference in the trajectories of the black holes between two levels than with the constraint violations. To quantify the trajectory difference between levels $i$ and $j$, we define
\begin{equation}
    \mathcal{D}^{(i,j)} \equiv \int_{t_1}^{t_2} \left\lVert \mathbf{x}_A^{(i)}(t)-\mathbf{x}_A^{(j)}(t) \right\rVert\,dt,
    \label{eq:integrated_trajectory_difference}
\end{equation}
where $\mathbf{x}_A^{(i)}(t)$ is the apparent-horizon coordinate center of black hole A at level $i$ \cite{sxs25_160}, $\lVert\cdot\rVert$ is the Euclidean norm, and $t_1$ and $t_2$ are the start and end times of the comparison. We plot $\mathcal{D}^{(i,j)}$ on the $x$-axis in Figures~\ref{fig_waveform_difference_combined} and~\ref{fig_waveform_difference_trajectory_correlation}. It is intended to capture the effect of the black holes following different coordinate trajectories at different levels. Another commonly used quantity that captures related effects is the phase difference between waveforms from two resolutions. We prefer $\mathcal{D}$ because it can be computed without waveform extraction and the errors associated with it.

\begin{table}[!t]
    \captionsetup{position=bottom}
    \centering
    {\renewcommand{\arraystretch}{1.25}%
    \setlength{\tabcolsep}{4pt}%
    \begin{tabular}{l|c|c|c|c}
         Levels & $\mathcal{D}_{\mathrm{Old}}$ & $\Delta_{\mathrm{Old}}$ & $\mathcal{D}_{\mathrm{New}}$ & $\Delta_{\mathrm{New}}$ \\
         \hline
         1 vs 6 & $22.9$ & $3.07 \times 10^{-6}$ & $3.91$ & $9.74\times 10^{-8}$ \\
         2 vs 6 & $39.4$ & $9.81 \times 10^{-6}$ & $5.20$ & $1.68\times 10^{-7}$ \\
         3 vs 6 & $7.03$ & $3.07 \times 10^{-7}$ & $0.964$ & $5.76\times 10^{-9}$ \\
         4 vs 6 & $0.304$ & $5.51 \times 10^{-10}$ & $0.0229$ & $4.50\times 10^{-12}$ \\
         5 vs 6 & $1.29$ & $1.02 \times 10^{-8}$ & $0.0532$ & $1.81\times 10^{-11}$ \\
         \hline
    \end{tabular}
    }
    \caption{Finite-radius waveform difference $\Delta$ and integrated trajectory difference $\mathcal{D}$, defined in Eqs.~\eqref{eq_waveform_difference} and~\eqref{eq:integrated_trajectory_difference}, for the Old and New simulations. No waveform alignment or optimization was performed when computing these differences. There is a very strong correlation between $\mathcal{D}$ and $\Delta$, as shown in Figure~\ref{fig_waveform_difference_combined}.}
    \label{tab_waveform_difference}
\end{table}

\begin{figure*}[!tp]
    \centering
    \includegraphics[width=0.98\linewidth]{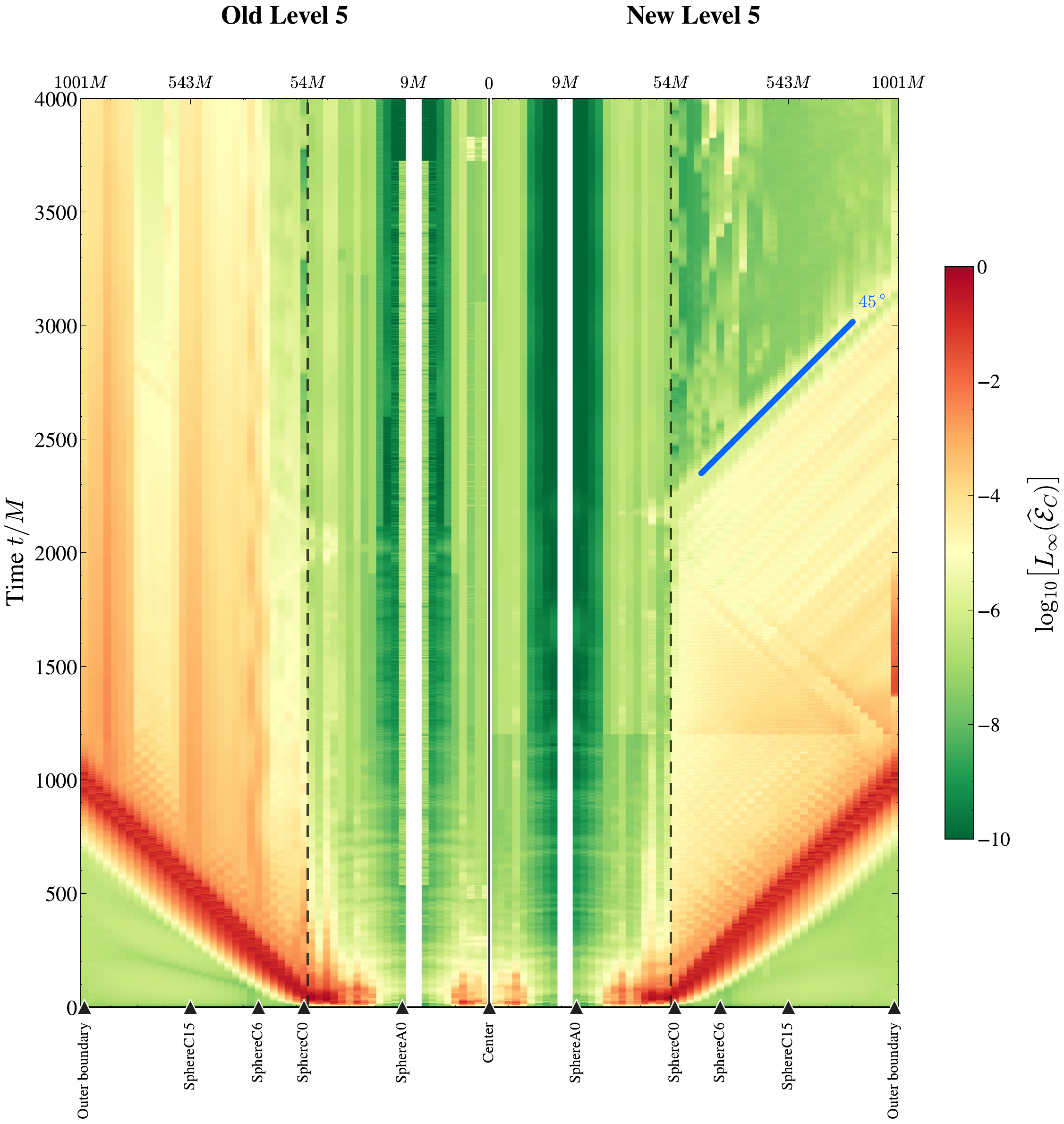}
    \caption{\label{fig_constraint_heatmaps_main_set1} Normalized constraint violations for the Old (left) and New (right) level 5 simulations. Both halves show the subdomains along the black-hole-A side of the line through the apparent-horizon coordinate centers of the two black holes. The two halves are arranged as mirror images, with the binary center toward the middle and the outer boundaries at the two edges. The vertical axis shows simulation time, and the color shows $\log_{10}[L_\infty(\widehat{\mathcal{E}}_{C})]$ over each subdomain. The bottom markers identify selected subdomains, while the top axis gives approximate distances from the binary center at $t=0$, increasing outward on both halves. The inner region up to SphereC0 is zoomed in, while the SphereC shells have approximately equal radial widths of $30M$ and are shown on a common linear scale. The finite-radius waveforms were extracted at $257M$, which lies in SphereC6. The white gaps mark the excised black-hole interiors, and the dashed lines mark the inner boundary of SphereC0. The bright red diagonal features show the outgoing junk-radiation pulse. Junk radiation travels at the speed of light and therefore appears at approximately $45^\circ$ with our chosen scales in the SphereC region, as indicated by the blue guide. After the junk radiation and its reflections have passed, the New simulation has smaller constraint violations in the outer subdomains. Also, in SphereA0 and the nearby subdomains, the New simulation is uniformly dark green (low constraint violations), while the Old simulation is not because of CFL violations.}
\end{figure*}

The main reason for this huge improvement of ``New'' compared to ``Old'' is that we are using a spatially uniform AMR tolerance for the whole computational domain in the New simulations. In particular, this means that the subdomains connecting the two black holes, colored pale gold in Figure~\ref{fig_SpEC_subdomains}, have more resolution for ``New''. Figure~\ref{fig_waveform_difference_combined} shows the waveform differences for the Old and New simulations together on a single plot for a direct comparison. We can see that, despite the difference in the numerical settings, the New and Old simulations, which started from the same initial data, lie nearly on the same line. A single fit describes both sets very well, with $\Delta$ scaling approximately as $\mathcal{D}^2$ during the inspiral part studied here. Because we have more resolution in the subdomains connecting the two black holes in the New simulations, the trajectories drift less across levels, leading to much smaller waveform differences.

We have also done separate simulations where we did not add more resolution and constraint damping in the outer subdomains, but added more resolution in the subdomains connecting the two black holes. The results were the same: we got two orders of magnitude improvement in the waveform differences. This highlights the fact that during the inspiral part of the simulation trajectory drift is the main factor that controls the waveform difference and not other factors like the waveform extraction errors.

We emphasize that waveform difference measures the similarity between two waveforms without any reference to the underlying physics. A small waveform difference means very little unless we also have convergence and small constraint violations. For example, an error in the implementation of the evolution equations could still produce waveform differences that converge with resolution. Another limitation of waveform difference is that the error in the less accurate waveform dominates the difference. It is possible that the level 6 waveform is much more accurate than suggested by the waveform difference between levels 5 and 6, because that difference is dominated by errors in the level 5 waveform. But we cannot make such claims without the waveform from a level 7 simulation.

For all of our BBH simulations, the waveform difference is dominated by the waveform's $(2,2)$ mode. This is especially true for the inspiral part of the system we are analyzing. If we neglect all the higher modes, the waveform difference will hardly change. While the overall waveform difference for the New simulations improved by two orders of magnitude, the higher-order waveform modes show little improvement at the higher resolutions. We can see this in Figures~\ref{fig_waveform_difference_main} and~\ref{fig_waveform_difference_set1}, where the $\ell=5$ and $\ell=6$ differences show little improvement. We suspect this is mostly caused by the issues discussed later in Section~\ref{sec:What limits our accuracy?}, which prevent these higher modes from being properly resolved. Note that we are only considering the inspiral portion of an equal-mass, non-spinning simulation far from merger. Close to the merger, these higher modes become much larger and are therefore better resolved.

Figure~\ref{fig_constraint_heatmaps_main_set1} compares the normalized constraint violations in the Old and New level 5 simulations across the computational domain. In the New simulation, the constraints reach their post-junk level in SphereC0 around $t=2000M$, whereas in SphereC29, located roughly $1000M$ from the center, they do so closer to $t=3000M$.

\subsection{Computational Cost of Improvements}
\label{ssec:Computational Cost of Improvements}
Because the improved waveform differences come from adding more grid points to the computational domain, it is important to ask how much this increases the computational cost. There are two possible reasons that can increase the cost. The first is that the ODE solver might need to take smaller time steps after we add more points because
of the CFL limit. The second is that each ODE step might take more wall time because there are more grid points to evolve. For these simulations, the first effect was not important: the ODE step size is limited by the subdomains closest to the black hole horizons, which we did not change. Therefore, the extra resolution in the connecting subdomains did not decrease the time step.

The extra grid points increase the total work per step, so each step becomes somewhat more expensive. However, the increase in wall time per step is modest. SpEC distributes work across cores in chunks of subdomains. Large subdomains get their own cores, while smaller subdomains are grouped together on a core. The goal is to give each core roughly the same amount of work. However, the work in a single subdomain is not split across multiple cores. For much of the simulation, the core with the largest subdomain takes the longest, while the other cores finish early and wait. The subdomains connecting the two black holes have far fewer grid points than the subdomains near the horizons. Adding resolution to these connecting subdomains, therefore, uses cores that would otherwise have spent more time waiting. In this sense, the improvement partly comes from using SpEC's parallelization more efficiently, so the wall-time increase is much smaller than one might expect from the added grid points alone.

We expect the same argument to hold for other BBH systems, but we have not run a sufficiently large set of simulations with these improvements to quantify the overall cost increase. The key point is that the step size is not reduced, and the increase in cost per step is small enough that the improvements in constraint violations and waveform differences remain worthwhile.

\section{What limits our accuracy?}
\label{sec:What limits our accuracy?}
Having looked at some improvements, we will now try to understand what prevents us from doing even more accurate simulations in SpEC. In other words, can we keep on increasing the resolution and expect the accuracy to keep on improving, provided computational resources are not a concern? As one might have guessed, the answer is no, and we go through some reasons for this below.

\subsection{Truncation errors and floating-point errors}
\label{ssec:Truncation errors and floating-point errors}
Because we use 64-bit floating-point numbers, there is a fundamental limit to how accurate we can be. Each arithmetic operation can introduce roundoff error of order $\epsilon_{\mathrm{mach}}\sim 10^{-16}$. In practice, each time step in SpEC requires a series of complex operations, which means that the cumulative effect of the floating-point errors is usually much larger.

\begin{figure}
    \includegraphics[width=\linewidth]{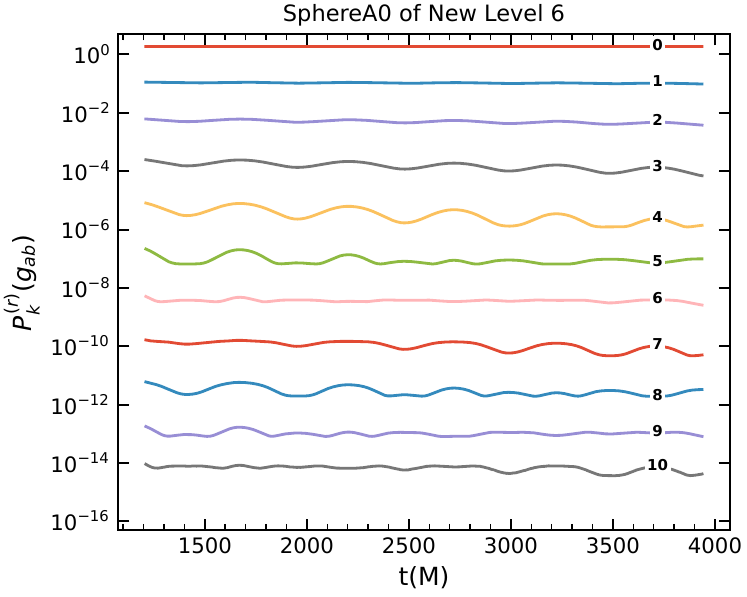}
    \caption{\label{fig_L16_set1_L6_PS_psi_SphereA0_0} Radial power monitors $P^{(r)}_k(g_{ab})$ versus time in SphereA0 for a New level 6 simulation.}
\end{figure}

\begin{figure}
    \includegraphics[width=\linewidth]{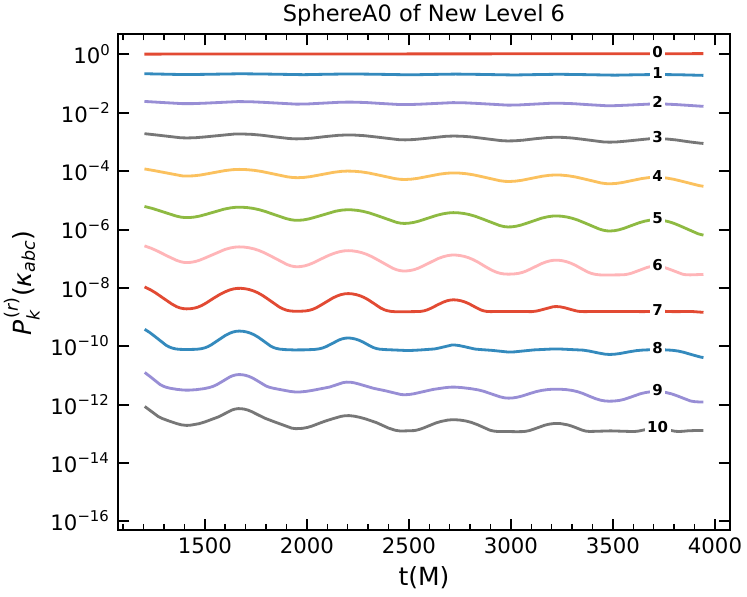}
    \caption{\label{fig_L16_set1_L6_PS_kappa_SphereA0_0} Radial power monitors $P^{(r)}_k(\kappa_{abc})$ versus time in SphereA0 for a New level 6 simulation.}
\end{figure}

These floating-point errors are usually small enough to be ignored, but there are subdomains in SpEC simulations where this becomes an issue. For example, in Figure~\ref{fig_L16_set1_L6_PS_psi_SphereA0_0}, the truncation error of the variable $g$ in the subdomain SphereA0 is around $10^{-14}$, which is close to the finite-precision noise floor, and we have reached this by using only 11 points in the radial direction. This is important because it indicates that we cannot just keep on adding more points and expect the simulation to get more and more accurate. Once our truncation error is comparable to this finite-precision noise floor, adding more points does not improve the accuracy but makes the CFL limit more restrictive. Not all subdomains have truncation errors comparable to the finite-precision noise floor because, in addition to the truncation error, the SpEC AMR algorithm also considers things like the constraint violations and the accuracy required to track the apparent horizon (Section~\ref{ssec:AMR tolerance and definition of levels}). In particular, some subdomains connecting the two black holes converge much more slowly, so we can keep adding more points in those subdomains without hitting the finite-precision limit. As we increase the levels further, we want the AMR to be more selective about the subdomains where it adds more points, which is a topic of future work.

From Figure~\ref{fig_L16_set1_L6_PS_kappa_SphereA0_0}, we can also see that the truncation error in the variable $\kappa$ is only $10^{-13}$, which is an order of magnitude larger than that of $g$. For our simulations, this is true in general: the truncation error in the variable $\kappa$ is always almost an order of magnitude larger than the truncation error in the variable $g$, reflecting the general property of loss of precision that happens on taking derivatives.

\begin{figure}
    \includegraphics[width=\linewidth]{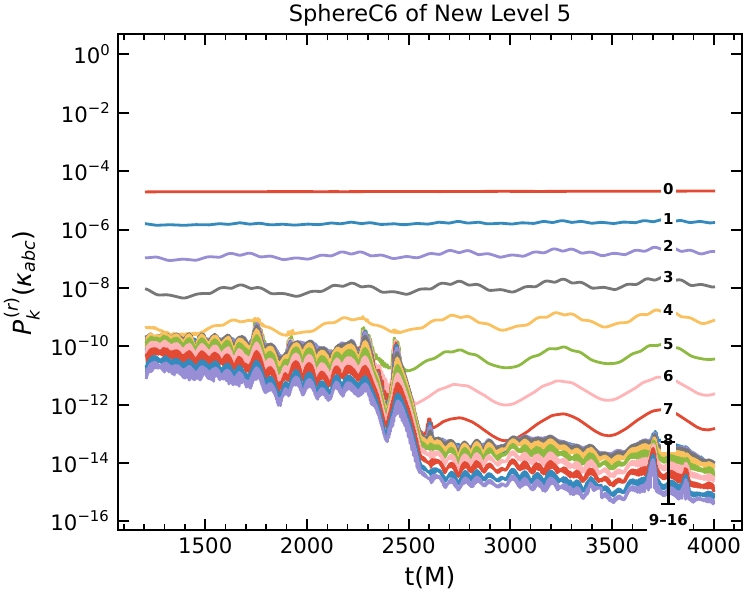}
    \caption{\label{fig_L16_set1_L5_PS_kappa_SphereC6_0} Radial power monitors $P^{(r)}_k(\kappa_{abc})$ versus time in SphereC6 for a New level 5 simulation.}
\end{figure}

\begin{figure}[!t]
    \includegraphics[width=\linewidth]{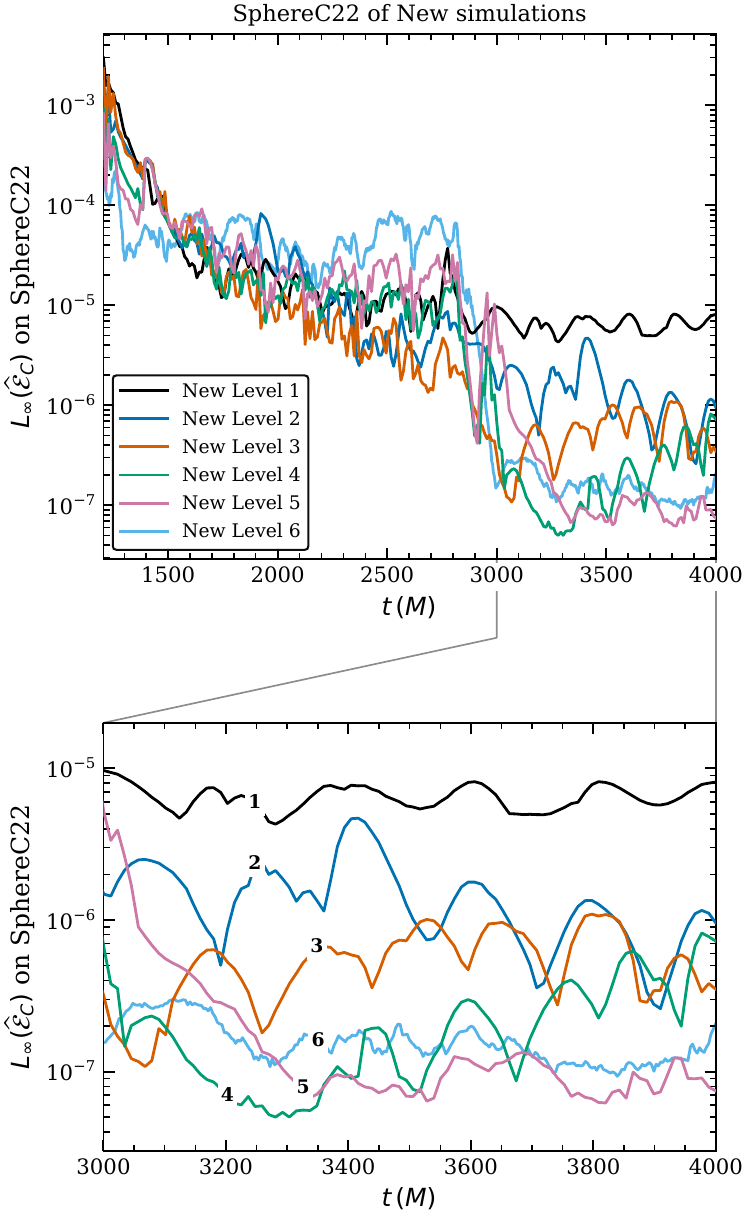}
    \caption{\label{fig_L16_set1_SphereC22_Linf_NormalizedGhCe} The $L_\infty$ norm of the normalized constraint energy density, $L_\infty(\widehat{\mathcal{E}}_{C})$, in SphereC22 ($r\approx760M$) versus time for New simulations at levels 1--6. The lower panel shows a zoom of $3000M\leq t\leq4000M$, with numbers identifying the resolution levels.}
\end{figure}

\subsection{Difference in scales of the evolution variables}
\label{ssec:Difference in scales of the evolution variables}
Far from the black holes, the spacetime metric $g_{ab}$ looks like a perturbation to the Minkowski metric,
\begin{equation}
    g_{ab} = \eta_{ab} + \delta g_{ab}.
\end{equation}
The dominant contributions roughly scale as
\begin{equation}
    \delta g_{ab}\sim\frac{M}{r},
    \qquad
    \kappa_{abc}\sim\frac{M}{r^2}.
\end{equation}
This means that the diagonal elements of the $g_{ab}$ are of order unity while other components are much smaller in magnitude. On the other hand, all the components of the derivative of the spacetime metric are much smaller. All these variables are used during the evolution in SpEC and are tightly coupled to each other, either by the way they appear in the GH formalism or because of things like constraint damping.

To understand why this matters, consider the values of these variables in SphereC6, located at approximately $r=250M$ in the computational domain. At this radius, $g_{00}$ is of order unity, so its spectral derivative can have a roundoff error of order $10^{-16}$. When we compute the three-index constraint violation (Eq.~\eqref{eq_constraint_3}) using this derivative, it inherits an error of the same order, which means that our three-index constraint will not fall below a finite-precision floor of order $10^{-16}$ even if we start with the analytical solution.

One might argue that this error is only present in the computation of the constraint violations, and the fields would still be accurate. However, when we add the constraint-damping terms in Eq.~\eqref{eq_full_GH_system}, errors in these terms enter the time derivatives of the evolution variables. This coupling means that the evolution variables have a constant source of absolute error, irrespective of the actual field values. The fields themselves fall rapidly as we go away from the center; therefore, the effect of this constant source of error is that the relative error in the fields grows as we go further away from the center. This can be seen in the spectral coefficients of the fields in the outer subdomains.
For example, in Figure~\ref{fig_L16_set1_L5_PS_kappa_SphereC6_0}, we can see that the highest radial power monitors have a noise floor around $10^{-15}$, and we have no convergence beyond that point, even though the scale of $\kappa_{abc}$ in this subdomain is about $10^{-5}$ ($\approx \frac{1}{250^2}$).
The direct implication of this noise floor is that the {\it relative} truncation error grows as we move further away from the center, and we cannot improve it by adding more points. Note that all this only becomes relevant once the residual junk radiation has either left the domain or has been damped away, which happens around $t=2500M$ in Figure~\ref{fig_L16_set1_L5_PS_kappa_SphereC6_0}. This also tells us that there is a sweet spot for how much constraint damping we can do because more constraint damping will only help up to a certain extent, after which the errors from the three-index constraints will start dominating. Unfortunately, this sweet spot depends on the resolution of the grid (which controls the error in the derivative) and the value of the fields, which change with time. Ideally, we want to have an algorithm that controls the amount of constraint damping based on these considerations, but a more detailed study of the problem would be required before implementing it.

While having a better algorithm for the constraint damping is an easy first step, it does not solve the more fundamental problem of having very different scales in the evolution variables. Ideally, we want to scale our fundamental variables to have similar magnitudes in the region of interest, but it is unclear how to do that in the GH system.

Another possibility is to make the spacetime metric minus the Minkowski metric one of our fundamental variables, $\bar{g}_{ab} = g_{ab} - \eta_{ab}$. This will not entirely solve the problem because $\bar{g}$ and $\kappa$ still fall off at different rates, but it could improve the situation by a few orders of magnitude because it removes the constant Minkowski contribution from the diagonal metric components. Consider Figure~\ref{fig_L16_set1_L5_PS_psi_SphereC6_0}, which shows the power spectrum of $g_{ab}$. Evolving $g_{ab} - \eta_{ab}$ instead would remove the dominant constant contribution to $P^{(r)}_{0}(g_{ab})$, which could allow us to resolve the remaining field more accurately. Note that $P^{(r)}_{1}(g_{ab})$ in Figure~\ref{fig_L16_set1_L5_PS_psi_SphereC6_0} is still a factor of 10 larger than $P^{(r)}_{0}(\kappa_{abc})$ in Figure~\ref{fig_L16_set1_L5_PS_kappa_SphereC6_0}. Thus, this change would reduce, but not eliminate, the four-orders-of-magnitude difference between the variables. We have performed single-black-hole simulations using the Minkowski-subtracted metric as the fundamental variable and observed improved constraints in the outer subdomains. Testing the same approach for a BBH system is a topic for future work.

We have already noted that the constraint violations do not converge after a certain resolution in the outer subdomains. At the same time, they continue to converge in the inner subdomains (Figure~\ref{fig_L15_ode_fix_SphereA0_Linf_GhCe.pdf}). We suspect the floating-point noise discussed above is one of the reasons for this. The fields in the outer domains are much smoother than those in the inner subdomains; therefore, they converge much faster, and we quickly hit the noise from calculating the constraint violations. For the simulation shown in Figure~\ref{fig_L16_set1_SphereC6_Linf_NormalizedGhCe}, we reach this noise floor by level 4, and adding points beyond that does not improve the constraint, as is the case for levels 5 and 6. We should mention that the large increase in the constraint violations for levels 5 and 6 in the figure is mostly caused by the glitches in the AMR (Appendix~\ref{app:Glitches due to grid changes}). This noise floor increases as we move further and further away from the center, which is why we see even less convergence in the subdomains that are closer to the outer boundary, like SphereC22, as shown in Figure~\ref{fig_L16_set1_SphereC22_Linf_NormalizedGhCe}. Once we switch to CCE, we will be able to extract gravitational waveforms using data from smaller radii, helping with this issue to some extent.

\section{Conclusions}
\label{sec:Conclusions}

In this paper, we studied the accuracy of SpEC BBH simulations during the inspiral. We found that the constraint convergence near the black holes was limited by the ODE tolerance, while reducing the constraint violations in the wave extraction region required both more resolution and more constraint damping. We also found that increasing the resolution of the subdomains connecting the two black holes made the black-hole trajectories more stable across different levels. With these changes, the finite-radius waveform differences improved by around two orders of magnitude. This last result is particularly relevant for other numerical-relativity codes because it is common practice to use very high resolution near the horizons but lower resolution in the region connecting the two black holes. We showed that the limiting factor for the waveform differences was the dephasing of the orbits between different resolutions, not the accuracy of gravitational-wave extraction. Therefore, allocating more resolution to the region connecting the two black holes, even at the cost of slightly lower resolution near the horizons, may be more effective at reducing waveform differences.

These improvements did not reduce the ODE step size in the simulations studied here because the additional grid points were added mainly in subdomains that were not limiting the time step. They do increase the work per step and, therefore, the wall time per step, but the increase is modest. The reason is that much of the extra work is assigned to cores that would otherwise spend more time waiting. This means that the accuracy of SpEC inspiral simulations can be improved substantially without a comparable increase in computational cost.

We also found that convergence in the outer subdomains at higher levels is affected by floating-point errors and AMR-related glitches. The improvement of around two orders of magnitude in the finite-radius waveform differences is very encouraging, given estimates of the waveform accuracy required for LISA and other planned detectors \cite{sxs25_83,sxs25_84}. Our results suggest that the changes presented here might allow us to reach these accuracy requirements. The next step is to run longer, end-to-end simulations and check this directly. We also need to check to what extent the improvements developed here carry over to more challenging cases, such as with high spins or high mass ratios.

\begin{acknowledgments}
    We thank Guillermo Lara for his detailed review of an early version of the manuscript. OpenAI Codex was used in writing some of the plotting scripts.
    This work was supported in part by the Sherman
    Fairchild Foundation, by NSF Grants PHY-2407742, PHY-2308615, PHY-2607486, and
    OAC-2513338 at Cornell, and
    NASA award 80NSSC26K0340; and by NSF Grants
    PHY-2309211, PHY-2309231, PHY-2607927, and OAC-2513339, and
    NASA award 80NSSC26K0340 at Caltech.
    Computations were performed at Caltech using the Resnick HPC Center.
\end{acknowledgments}

\appendix

\section{Generalized Harmonic System}
\label{app:Generalized Harmonic System}

Full evolution equations in the GH formalism are:
\begin{widetext}
\begin{align}
    \begin{aligned}
    \partial_t g_{a b}-\left(1+\gamma_1\right) N^k \partial_k g_{a b}
    &= -N \Pi_{a b}-\gamma_1 N^i \Phi_{i a b}, \\
    \partial_t \Pi_{a b}-N^k \partial_k \Pi_{a b}+N \gamma^{k i} \partial_k \Phi_{i a b}-\gamma_1 \gamma_2 N^k \partial_k g_{a b}
    &= 2 N g^{c d}\left(\gamma^{i j} \Phi_{i c a} \Phi_{j d b}-\Pi_{c a} \Pi_{d b}-g^{e f} \Gamma_{a c e} \Gamma_{b d f}\right) \\
    &\quad -2 N \nabla_{(a} H_{b)}-\frac{1}{2} N t^c t^d \Pi_{c d} \Pi_{a b}-N t^c \Pi_{c i} \gamma^{i j} \Phi_{j a b} \\
    &\quad +N \gamma_0\left[2 \delta^c{ }_{(a} t_{b)}-g_{a b} t^c\right]\left(H_c+\Gamma_c\right)-\gamma_1 \gamma_2 N^i \Phi_{i a b}, \\
    \partial_t \Phi_{i a b}-N^k \partial_k \Phi_{i a b}+N \partial_i \Pi_{a b}-N \gamma_2 \partial_i g_{a b}
    &= \frac{1}{2} N t^c t^d \Phi_{i c d} \Pi_{a b}+N \gamma^{j k} t^c \Phi_{i j c} \Phi_{k a b}-N \gamma_2 \Phi_{i a b} .
    \end{aligned}
\label{eq_full_GH_system}
\end{align}
\end{widetext}
The parameters $\gamma_0$, $\gamma_1$, and $\gamma_2$ are used for adjusting characteristic speeds and for constraint damping \cite{sxs25_153}; see Section~\ref{ssec:Constraint violations and Constraint damping}. Their contributions are proportional to the constraints and vanish when the constraints are exactly satisfied.

\section{Glitches due to resolution changes}
\label{app:Glitches due to grid changes}
In addition to the junk radiation, an unresolved but understood source of error, the right half of Figure~\ref{fig_constraint_heatmaps_main_set1} also shows error spikes in some outer subdomains, such as SphereC6 and neighboring subdomains around $t\approx3700M$. These glitches occur at irregular times and usually do not propagate outward like the junk radiation. We strongly suspect that they are caused by AMR resolution changes. Each AMR change introduces a small discontinuity in time where the dynamical variables are suddenly represented by a different number of basis functions. Ideally, these discontinuities should be below the truncation error, but because they occur at irregular times, they introduce a small source of noise into the simulation.

To test whether the glitches are indeed caused by AMR,
we did some simulations where the AMR was turned off, with the grid resolution manually set to be similar to what AMR would have chosen. In this case, we did not observe any glitches. In most cases, the glitches are not really relevant because they are relatively rare and only affect the outer subdomains. We mention them here because we see them in heatmap figures, for example, in the subdomain SphereC29 in the right half of Figure~\ref{fig_constraint_heatmaps_main_set1} and some constraint violation plots like Figure~\ref{fig_L16_set1_SphereC6_Linf_NormalizedGhCe}.

\onecolumngrid
\section{Initial and relaxed parameters}
\label{app:Parameters of the systems}

\begin{table}[!h]
    \caption{Initial-data and relaxed parameters for the simulations used in this paper. The ID row gives the initial-data solver output, while the New and Old/ODE fix rows give the relaxed parameters measured after PBandJ. The simulation labels are defined in Section~\ref{sec:Improvements to SpEC}.}
    \begin{ruledtabular}
    \begin{tabular}{lccccc}
        Case & Orbital eccentricity & $M_A$ & $M_B$ & $\chi_A$ & $\chi_B$ \\
        \hline
        ID & -- & $0.4999999739$ & $0.4999999688$ &
        $\begin{pmatrix}
            -9.194\times10^{-11} \\
            -9.288\times10^{-12} \\
            -2.138\times10^{-8}
        \end{pmatrix}$ &
        $\begin{pmatrix}
            2.648\times10^{-11} \\
            3.037\times10^{-11} \\
            -2.545\times10^{-8}
        \end{pmatrix}$ \\
        New & $0.0515847$ & $0.4999998464$ & $0.4999998413$ &
        $\begin{pmatrix}
            6.140\times10^{-14} \\
            2.002\times10^{-15} \\
            -2.651\times10^{-7}
        \end{pmatrix}$ &
        $\begin{pmatrix}
            1.894\times10^{-14} \\
            -3.425\times10^{-16} \\
            -2.612\times10^{-7}
        \end{pmatrix}$ \\
        Old/ODE fix & $0.0540567$ & $0.5000004262$ & $0.5000004209$ &
        $\begin{pmatrix}
            2.670\times10^{-12} \\
            4.792\times10^{-12} \\
            -2.027\times10^{-6}
        \end{pmatrix}$ &
        $\begin{pmatrix}
            -2.068\times10^{-12} \\
            -4.602\times10^{-12} \\
            -2.026\times10^{-6}
        \end{pmatrix}$
    \end{tabular}
    \end{ruledtabular}
    \label{tab_system_parameters}
\end{table}

\twocolumngrid

\normalem %
\bibliography{main}%

\end{document}